\documentclass[12pt]{article} 
\usepackage[sectionbib]{natbib}
\usepackage{array,epsfig,fancyhdr,rotating}
\usepackage[]{hyperref}  
\usepackage{sectsty, secdot}
\sectionfont{\fontsize{12}{14pt plus.8pt minus .6pt}\selectfont}
\renewcommand{\theequation}{\thesection\arabic{equation}}
\subsectionfont{\fontsize{12}{14pt plus.8pt minus .6pt}\selectfont}

\usepackage{amsmath}
\usepackage{amssymb}
\usepackage{amsfonts}
\usepackage{multirow}
\usepackage{amsthm}
\usepackage{bm, bbm, color}

\newtheorem{theorem}{Theorem}

\newtheorem{corollary}{Corollary}

\theoremstyle{definition}

\newtheorem{remark}{Remark}
\newtheorem{assumption}{Assumption}
\usepackage{subfigure}
\usepackage{bm,bbm}
\usepackage{multirow}
\usepackage{array}
\usepackage{warpcol}
\usepackage{booktabs}
\newcommand{\tabincell}[2]{\begin{tabular}{@{}#1@{}}#2\end{tabular}}
\usepackage{setspace}

\usepackage{xr}
\usepackage{xcolor}

\newif\ifred
\redfalse   

\newcommand{\revise}[1]{%
	\ifred
	\textcolor{red}{#1}%
	\else
	#1%
	\fi
}

\begin{document}


\renewcommand{\baselinestretch}{2}

\markright{ \hbox{\footnotesize\rm Statistica Sinica
}\hfill\\[-13pt]
\hbox{\footnotesize\rm
}\hfill }

\markboth{\hfill{\footnotesize\rm YUXIN TAO, HUAN GONG AND DONG LI} \hfill}
{\hfill {\footnotesize\rm ASYMMETRIC GARCH WITHOUT MOMENT CONDITIONS} \hfill}

\renewcommand{\thefootnote}{}
$\ $\par


\fontsize{12}{14pt plus.8pt minus .6pt}\selectfont \vspace{0.2pc}
\centerline{\large\bf ASYMMETRIC GARCH MODELLING WITHOUT }
\vspace{1.8pt} 
\centerline{\large\bf MOMENT CONDITIONS}
\vspace{.12cm} 
\centerline{Yuxin Tao, Huan Gong\textsuperscript{*} , Dong Li} 
\renewcommand{\thefootnote}{*}
\begin{footnotetext}
	{Corresponding author.
		\textit{E-mail:} gonghuan@nudt.edu.cn}
\end{footnotetext}
\vspace{.08cm} 
\centerline{\it Southern University of Science and Technology, National University of}
\centerline{\it Defense Technology, Tsinghua University}
 \vspace{0cm} \fontsize{9}{11.5pt plus.8pt minus.6pt}\selectfont


\begin{quotation}
\noindent {\it Abstract:}
Heavy tails and stability are two persistent challenges in modelling financial time series, yet most existing approaches rely on finite-moment assumptions and pay insufficient attention to stability issues.
To bridge this gap, we propose an asymmetric GARCH model with standardized non-Gaussian stable innovations (sAGARCH), which accommodates infinite variance and even infinite mean.
We establish a comprehensive inference framework for both stationary and explosive cases, proving the strong consistency and asymptotic normality of the maximum likelihood estimator, including the tail index parameter.
We also discuss multiple estimators for the asymptotic variance.
Additionally, we propose a modified Kolmogorov-type test statistic for diagnostic checking, along with tests for strict stationarity and asymmetry.
Through Monte Carlo simulations with heavy-tailed innovations, we provide further insight into the finite-sample performance of the intercept estimator.
Empirical applications to stock returns further highlight the usefulness and merits of the proposed sAGARCH model.

\vspace{0pt}
\noindent {\it Key words and phrases:}
Heavy tails, Kolmogorov-type test, Maximum likelihood estimation, Nonstationarity, Stable distribution.
\par
\end{quotation}\par

\def\thefigure{\arabic{figure}}
\def\thetable{\arabic{table}}

\renewcommand{\theequation}{\thesection.\arabic{equation}}

\fontsize{12}{14pt plus.8pt minus .6pt}\selectfont

\section{Introduction}

\revise{Heavy-tailed phenomena are ubiquitous across scientific fields \citep{Harvey:2013, Nolan}, and have received sustained attention.
Financial time series are a typical example, as they often exhibit heavy tails, volatility clustering, and asymmetry.}
Since the seminal work of \cite{Engle:1982} and \cite{Bollerslev:1986}, the Generalized Autoregressive Conditional Heteroskedasticity (GARCH) model has become a benchmark for econometricians and financial practitioners \citep{Fan:Yao}, with numerous variants proposed for practical needs \citep{FZ:2019}.
\revise{
Although GARCH-type models are designed to capture several stylized features, they often struggle with tail thickness, and excess kurtosis may remain in the standardized residuals of fitted models \citep{bai}.
Thus, the commonly imposed finite fourth- or second-moment assumptions on the innovations can be too restrictive.
Student's $t$ and generalized Gaussian innovations are useful alternatives, but they lack stability under addition; see, e.g., \cite{fama}, \cite{Samuelson}, \cite{Nolan}, and discussions in \cite{Li:2023}. }

\revise{To address this issue, we propose the use of stable distributions for innovations in volatility models.
Their application in finance dates back to \cite{Mandel} and \cite{fama}, who advocated stable laws for modelling unconditional asset returns, 
although early use was limited by the lack of closed-form densities.
With modern numerical methods, stable models have become computationally feasible.}
A common concern is that non-normal stable distributions may have infinite mean or variance, seemingly contradicting the finite sample moments observed in practice. 
However, as noted by \cite{Mandel1997}, sample moments need not converge asymptotically, so empirical finiteness does not imply finite theoretical moments.
Indeed, power-law models with divergent moments (tail indices below 2 or even 1) are well-documented in finance and physics \citep{UZ},
and infinite-mean models may provide useful statistical fits and financial risk insights \citep{ChenWang2025}.

The use of stable innovations in GARCH models was first proposed by \cite{McCulloch} and later extended by \cite{Liu:1995}, mainly from an applied perspective.
Stationarity conditions were studied by \cite{Panorska} and \cite{Mittnik}.
For estimation, \cite{Liu:1995} proposed the maximum likelihood estimation (MLE) approach, while \cite{chp2014} developed an indirect inference method using Student's $t$ innovations as auxiliary models.
On the other hand,
\cite{Hall:2003} analyzed heavy-tailed GARCH models with innovations having infinite fourth moments but finite variance.
\cite{Zhang2022} explored the least absolute deviation estimator for autoregressive models with heavy-tailed G-GARCH noise.
However, to the best of our knowledge,  the asymptotic properties of MLE and statistical inference for GARCH-type models with infinite-moment innovations remain unexplored in the literature.

Conditional asymmetry, or leverage effects, is another important feature of financial time series.
Recent works on asymmetric GARCH-type models include \cite{Li:2020}, \cite{Wang2022}, and \cite{Zhu2023}.
\revise{For stable innovations, asymmetry could be introduced through a nonzero skewness parameter $\beta$.
However, skew stable densities are computationally more complicated and may cause numerical instability.}
Moreover, as the stability exponent $\alpha$ approaches 2, stable distributions tend to symmetry, making $\beta$ meaningless and harder to estimate
(especially for $\alpha = 2$, which is the Gaussian case); see \cite{Nolan}.
\revise{We therefore model asymmetry through the volatility structure rather than through skewed stable innovations.}

Our major contributions are threefold.
\revise{First, we propose an asymmetric GARCH model with stable innovations, which accommodates excess kurtosis, volatility clustering, and leverage effects without imposing moment conditions.
We establish the strong consistency and asymptotic normality of the MLE under a unified framework covering both stationary and explosive cases, with $n^{-1/2}$ convergence rate except for the intercept in the explosive case. It differs from the findings of \cite{Hall:2003}.
We also provide two estimators of the Fisher information matrix and a universal estimator of the asymptotic variance that does not require stationarity.}


\revise{Second, we develop several technical tools for handling infinite-moment stable innovations in GARCH-type models.
These include refined asymptotic bounds for derivatives of log-stable densities in Appendix \ref{appC}, an identifiability argument for the stability parameter $\alpha$ in Lemma \ref{garch.lemma.new}, and an improved exponential convergence rate for the volatility recursion in Lemma \ref{lemma1}.
These results are essential for removing moment restrictions.}

\revise{Third, we propose inference procedures for diagnostic checking, strict stationarity, and asymmetry testing.
In particular, we develop a modified Kolmogorov-type diagnostic statistic using the transformation method of \cite{bai}.
Monte Carlo studies verify the finite-sample performance of the MLE and the proposed tests, and empirical applications to stock returns illustrate the practical advantages of the proposed model.}

The remainder of the paper is organized as follows.
Section \ref{section2} proposes the sAGARCH model and derives its stationarity conditions.
Section \ref{MLE} considers the asymptotic properties of the MLE, a universal estimator of its asymptotic variance, and computational issues.
Section \ref{test} develops tests for diagnostic checking, strict stationarity, and asymmetry.
Section \ref{simulation} reports the finite-sample performance of MLE and test statistics.
Section \ref{example} analyzes individual stock returns.
Section \ref{conclusion} concludes the paper.
All technical proofs are postponed to the Supplementary Material.

\section{The sAGARCH model: Stability and Explosivity} 
\label{section2}
\subsection{Stable distribution}
Stable distributions are defined by their characteristic functions, since their densities have no closed forms except for three special cases (i.e., Gaussian, Cauchy, and L\'{e}vy distributions).
A general stable distribution $\mathbf{S}(\alpha,\beta,\gamma,\delta)$  contains four parameters: $\alpha, \beta, \gamma$ and $\delta$,  representing stability, skewness, scale, and location, respectively. 
In this paper, we consider the standardized
symmetric stable
distribution $\mathbf{S}(\alpha,0,1,0)$, of which the characteristic function is
$\phi(s)=
\exp(-|s|^{\alpha})$, $ s\in \mathbb{R},$
where $\alpha\in(0, 2]$. Its density is
\begin{flalign}\label{density}
	f_{\alpha}(x)
	=\frac{1}{\pi}\int_0^\infty\exp(-s^{\alpha})\cos(sx)ds
	,\quad x\in \mathbb{R}.
\end{flalign}
It has no variance when $\alpha <2$ and no expectation when $\alpha \leq 1$.
Particularly, when $\alpha=1$, it is the standard Cauchy distribution, and when $\alpha=2$, it reduces to $\mathcal{N}(0, 2)$.
More useful properties on stable distribution can be found in Propositions \ref{prop.2.3}-\ref{prop.2.5} in the Appendix \ref{appC}.

\subsection{The sAGARCH model}
To better capture excess kurtosis, asymmetry and volatility clustering jointly, we define the first-order asymmetric GARCH model with standardized non-Gaussian
symmetric stable innovation (hereafter sAGARCH(1,1)) as
\begin{flalign}\label{sgarch}
	\left\{
	\begin{array}{l}
		y_t=\sigma_t \eta_t, \quad t \in \mathbb{Z}_{+}:=\{1, 2,...\},\\
\sigma_{t}^{2}=\omega+\phi_{+}\left(y_{t-1}^{+}\right)^{2}+\phi_{-}\left(y_{t-1}^{-}\right)^{2}+\psi \sigma_{t-1}^{2},
	\end{array} 
    \right.
\end{flalign}
with initial values $y_{0}$ and $\sigma_{0} \geq 0$, where $\omega>0$, $\phi_{+} > 0$, $\phi_{-} > 0$, $\psi > 0$, $x^{+}=\max \{x, 0\}$, $x^{-}=-\min \{x, 0\}$.
$\{\eta_t\}$ is a sequence of
i.i.d. standardized non-Gaussian symmetric stable random variables
with the stable exponent $\alpha\in(0, 2)$. 
Heavy-tailedness is modeled through stable innovation $\{\eta_t\}$, and asymmetry is captured by different parameters $\phi_+$ and $\phi_-$.

By \cite{BP:1992a, BP:1992b}, model (\ref{sgarch}) is strictly stationary \textit{if and only if}
the top Lyapunov exponent
\begin{flalign} \label{Lyapunov}
	\gamma_{\alpha}:= E\log a(\eta_1)<0, \quad
	\mbox{where} ~ a(x) =\phi_{+}(x^{+})^{2}+\phi_{-}(x^{-})^{2}+\psi.
\end{flalign}
\revise{The plots of the stable densities and the strict stationarity regions under different scenarios are provided in the Appendix \ref{appA0}.}

\section{Maximum Likelihood Estimation}\label{MLE}
Let $\vartheta = (\phi_+, \phi_-, \psi, \alpha)'$, and $\theta=(\omega, \vartheta')'=(\omega, \phi_+, \phi_-, \psi, \alpha)' \in \mathbb{R}_+^4 \times (0, 2)$, where $\mathbb{R}_+=(0, \infty)$.
Suppose that the observations $\{y_0, y_1,...,y_n\}$ are from model (\ref{sgarch}) with
true parameter $\theta_0=(\omega_0, \vartheta_0')'=(\omega_0, \phi_{0+}, \phi_{0-}, \psi_0,  \alpha_0)'$.
Then the (conditional) log-likelihood function is defined as
\begin{equation} \label{Li-five}
	\widetilde{L}_n(\theta)=\sum_{t=1}^n  \widetilde{\ell}_t(\theta), \quad
	\widetilde{\ell}_t(\theta)=-\log \widetilde{\sigma}_{t}(\theta)+\log f_\alpha\Big(\frac{y_t}{\widetilde{\sigma}_{t}(\theta)}\Big),
\end{equation}
where $f_\alpha(x)$ is defined in (\ref{density}), and
$\widetilde{\sigma}_{t}^{2}(\theta)=\omega+\phi_{+}(y_{t-1}^{+})^{2}+\phi_{-}(y_{t-1}^{-})^{2}+\psi \widetilde{\sigma}_{t-1}^{2}(\theta)$, $t \geq 1,$
with initial values $(y_0, \widetilde{\sigma}_0 (\theta))\equiv (0,0)$.
The MLE of $\theta_0$ is
\begin{flalign*}
	\widehat{\theta}_n:=(\widehat{\omega}_n, \widehat{\vartheta}_n) :=(\widehat{\omega}_n, \widehat{\phi}_{n+}, \widehat{\phi}_{n-}, \widehat{\psi}_n,  \widehat{\alpha}_n)' = \arg\max_{\theta\in\Theta} \widetilde{L}_n(\theta),
\end{flalign*}
where the parameter space $\Theta$ is a subset of $\mathbb{R}_+^4\times(0, 2)$.

\subsection{Consistency and asymptotic normality}
To study the asymptotics of $\widehat{\theta}_n$,  we require the following assumptions.
\begin{assumption}\label{asm1}
	$\{\eta_t\}$ is a sequence of i.i.d. standardized non-Gaussian symmetric stable random variables
	with the density function $f_{\alpha_0}(x)$.
\end{assumption}
\begin{assumption}\label{asm2}
	The parameter space $\Theta$ is compact and $\theta_0\in\Theta$.
\end{assumption}
\begin{assumption}\label{asm3}
	The true parameter $\theta_0$ is an interior point of $\Theta$.
\end{assumption}

We are now ready to present our main results. Theorem~\ref{theorem2.1} establishes the results for the stationary case, and Theorem~\ref{theorem2.2} addresses the explosive case.
To study the asymptotic properties of $\widehat{\theta}_n$ for the stationary case, we define the theoretical (conditional) log-likelihood function as
$L_n(\theta)=\sum_{t=1}^n \ell_t(\theta)$ with
$\ell_t(\theta)=-\log \sigma_{t}(\theta)+\log f_\alpha(y_t/\sigma_{t}(\theta))$,
$\sigma_{t}^{2}(\theta)=\omega+\phi_{+}(y_{t-1}^{+})^{2}+\phi_{-}(y_{t-1}^{-})^{2}+\psi \sigma_{t-1}^{2}(\theta),  t \in \mathbb{Z}.$
Then when $0 < \psi<1$,
\begin{flalign} \label{garch.3.8}
	\sigma_{t}^{2}(\theta)= \sum_{j=0}^{\infty} \psi^j \{\omega+\phi_{+}(y_{t-1-j}^{+})^{2}+\phi_{-}(y_{t-1-j}^{-})^{2}\}, \quad  t \in \mathbb{Z}.
\end{flalign}
Since $\{y_t: t\in \mathbb{Z}\}$ is strictly stationary and ergodic, so is $\{\sigma_{t}^{2}(\theta): t\in \mathbb{Z}\}$ by the measurability of $\sigma_{t}^{2}(\theta)$ with respect to $\{y_t\}$.

\begin{theorem}\label{theorem2.1}
	Suppose that Assumptions \ref{asm1}--\ref{asm2} hold and $\gamma_{\alpha_0}<0$. \\
	\noindent $(\mathrm{i})$.
	For $\Theta$ such that $\forall\theta \in \Theta$, $\psi<1$, then $\widehat{\theta}_n \xrightarrow{a.s.} \theta_0$, as $n\rightarrow\infty$.\par
	\noindent $(\mathrm{ii})$. Further, 
    if Assumption \ref{asm3} holds, then
	$\sqrt{n}(\widehat{\theta}_n-\theta_0) \xrightarrow{d} \mathcal{N}(0, \Sigma^{-1})$, where `$\xrightarrow{d}$' stands for convergence in distribution and
	\begin{flalign*}
		\Sigma=E\Big\{\frac{\partial\ell_t(\theta_0)}{\partial\theta}
		\frac{\partial\ell_t(\theta_0)}{\partial\theta'}\Big\}
		=\Bigg(
		\begin{array}{c c}
			\Sigma_{\tilde{\theta} \tilde{\theta}^{\prime}} & \Sigma_{\tilde{\theta}  \alpha}' \\
			\Sigma_{\tilde{\theta}  \alpha} & \Sigma_{\alpha\alpha}
		\end{array}
		\Bigg)
	\end{flalign*}
	with $\tilde{\theta}=(\omega, \phi_+, \phi_-, \psi)'$ and
	\begin{flalign*}
		\quad&\Sigma_{\tilde{\theta} \tilde{\theta}^{\prime}}
		=\frac{1}{4}E\Big\{\frac{1}{\sigma_t^4(\theta_0)} \frac{\partial \sigma_t^2(\theta_0)}{\partial \tilde{\theta}} \frac{\partial \sigma_t^2(\theta_0)}{\partial \tilde{\theta}^{\prime}}\Big\}
		E\left[\Big\{1+\frac{\partial \log f_{\alpha_0}(\eta_t)}{\partial x}\eta_t\Big\}^2\right],\\
		&\Sigma_{\tilde{\theta}  \alpha}
		=-\frac{1}{2}E\Big\{\frac{1}{\sigma_t^2(\theta_0)} \frac{\partial \sigma_t^2(\theta_0)}{\partial \tilde{\theta}}\Big\}E\Big\{
		\frac{\partial \log f_{\alpha_0}(\eta_t)}{\partial x}
		\frac{\partial \log f_{\alpha_0}(\eta_t)}{\partial \alpha}\eta_t\Big\}, \\
		&\Sigma_{\alpha\alpha}
		=E\left[\Big\{\frac{\partial \log f_{\alpha_0}(\eta_t)}{\partial \alpha}\Big\}^2\right]. &&
	\end{flalign*}
\end{theorem}

\begin{theorem}\label{theorem2.2}
	Suppose that Assumptions \ref{asm1}--\ref{asm2} hold and $\gamma_{\alpha_0}>0$, then\par
	\noindent $(\mathrm{i})$. 
	$\widehat{\vartheta}_n \xrightarrow{a.s.} \vartheta_0$,  as $n\rightarrow\infty$.\\
	\noindent $(\mathrm{ii})$. 
	Further, if Assumption \ref{asm3} holds, then $\sqrt{n}\big(\widehat{\vartheta}_n-\vartheta_0\big) \xrightarrow{d} \mathcal{N}(0, \Upsilon^{-1})$, where
	$\Upsilon=\Bigg(
	\begin{array}{c c}
		\Upsilon_{\widetilde{\vartheta} {\widetilde{\vartheta}}^{\prime}} & \Upsilon_{\widetilde{\vartheta}  \alpha}' \\
		\Upsilon_{\widetilde{\vartheta}  \alpha} & \Upsilon_{\alpha\alpha}
	\end{array}
	\Bigg)$
	with $\tilde{\vartheta}=(\phi_+, \phi_-, \psi)'$ and
	\begin{flalign*}
		&\Upsilon_{\widetilde{\vartheta} {\widetilde{\vartheta}}^{\prime}}
		=\frac{1}{4}E(d_t d_t^{\prime}) E\left[\Big\{1+\frac{\partial \log f_{\alpha_0}(\eta_t)}{\partial x} \eta_t\Big\}^2\right],\\
		&\Upsilon_{\widetilde{\vartheta}  \alpha}
		=-\frac{1}{2} E(d_t) E \Big\{\frac{\partial \log f_{\alpha_0}(\eta_t)}{\partial x} \frac{\partial \log f_{\alpha_0}(\eta_t)}{\partial \alpha} \eta_t\Big\}, ~
		\Upsilon_{\alpha\alpha}
		=E\left[\Big\{\frac{\partial \log f_{\alpha_0}(\eta_t)}{\partial \alpha}\Big\}^2\right],\\
		&E(d_{t})=E(d_{t}^{\phi_{+}}, d_{t}^{\phi_{-}}, d_{t}^{\psi})'
		= \Big(\frac{1-\nu_{1+}}{\phi_{0+}(1-\nu_{1})}, \frac{1-\nu_{1-}}{\phi_{0-}(1-\nu_{1})}, \frac{\nu_{1}}{\psi_0(1-\nu_{1})}\Big)^\prime,    \\
		&E\{(d_{t}^{\phi_{+}})^2\} =\frac{ (1-2 \nu_{1+}+\nu_{2+} ) (1-\nu_{1} )+2 (\nu_{1+}-\nu_{2+} ) (1-\nu_{1+} )}{\phi_{0+}^{2} (1-\nu_{1} ) (1-\nu_{2} )},\\
		&E(d_{t}^{\phi_{+}} d_{t}^{\phi_{-}})= \frac{ (\nu_{1+}-\nu_{2+} ) (1-\nu_{1-} )+ (\nu_{1-}-\nu_{2-} ) (1-\nu_{1+} )}{\phi_{0+} \phi_{0-} (1-\nu_{1} ) (1-\nu_{2} )},\\
		&E(d_{t}^{\phi_{+}} d_{t}^{\psi})= \frac{\nu_{2} (1-\nu_{1+} )+\nu_{1+}-\nu_{2+}}{\psi_{0} \phi_{0+} (1-\nu_{2} ) (1-\nu_{1} )}, \quad 
		E\{(d_{t}^{\psi})^2\}= \frac{\nu_{2} (1+\nu_{1} )}{\psi_{0}^{2} (1-\nu_{2} ) (1-\nu_{1} )},
		&&
	\end{flalign*}
	and $E\{(d_{t}^{\phi_{-}})^2\}$ (or $E(d_{t}^{\phi_{-}} d_{t}^{\psi})$) is obtained by replacing $\phi_{0+}$ by $\phi_{0-}$ and $\nu_{i+}$ by $\nu_{i-}$ in $E\{(d_{t}^{\phi_{+}})^2\}$ (or $E(d_{t}^{\phi_{+}} d_{t}^{\psi})$),
	with 
	$\nu_{i}=E\left[\{ \psi_{0}/a_{0}(\eta_{t})\}^{i}\right]$, 
	$\nu_{i+}=E\left[\{\psi_{0}/(\phi_{0+}(\eta_{t}^{+})^2+\psi_{0}) \}^{i}\right]$, 
	$\nu_{i-}=E\left[\{\psi_{0}/(\phi_{0-}(\eta_{t}^{-})^2+\psi_{0}) \}^{i}\right]$, for $i=1,2$.
\end{theorem}

\begin{remark}
When $\alpha_0=1$, similar to \cite{Li:2023}, the exact value of the Fisher matrix $\Upsilon$ in the explosive case can be calculated. The derivation is presented in the Appendix \ref{appD} in the Supplementary Material.
\end{remark}

\begin{remark}
	Theorem \ref{theorem2.2} shows that the intercept $\omega_0$ is not estimable in the explosive case, due to the non-identifiability of $\omega$ in the limit of $L_n(\theta)/n$.
	This partial parameter unidentifiability was first observed in QMLE of nonstationary ARCH(1)  \citep{jra,jrb} and later in nonstationary GARCH models \citep{FZ:2012,FZ:2013}.
\end{remark}

\begin{remark}
	Theorems \ref{theorem2.1}-\ref{theorem2.2} exclude the critical case $\gamma_{\alpha_0}=0$. 
	To the best of our knowledge, the behavior of $y_t$ in this critical case remains largely unresolved in the literature.
	For example, although \cite{FZ:2012} claimed to address this issue, their additional Assumption A is difficult to verify under stable innovations. Our extensive simulations suggest that their assumption does not hold.
	While we could impose simpler assumptions on
	$y_t$ to obtain the consistency and asymptotic normality of $\widehat{\vartheta}_n$, such as $|y_t|/\rho^t \xrightarrow{a.s.} \infty$, $\rho>1$,  as $t\rightarrow\infty$, such conditions are typically unverifiable in practice.
	Thus, we leave this problem for future research.
\end{remark}

\begin{remark}
	As discussed in the Introduction, another way to model asymmetry is to use  $\eta_t\sim \mathbf{S}(\alpha,\beta,1,0)$ with skewness parameter  $\beta \in (-1,1)$.
	However, this poses substantial numerical challenges, as the density $f_{\alpha,\beta}$ has no closed form,  exhibits unreliable behavior near boundary cases, and is computationally expensive to evaluate repeatedly in optimization.
	In addition, as $\alpha\to 2$, stable laws approach the Gaussian law, and $\beta$ becomes less informative and harder to estimate reliably.
	Theoretically, extending to skew-stable GARCH(1,1) setting mainly requires: 
	(i) the asymptotic behavior of partial derivatives of $\log f_{\alpha,\beta}$ involving $\beta$, and (ii) the identifiability of the skewness parameter $\beta$.
	We leave this extension to future work.
\end{remark}

\subsection{Estimation of Fisher matrices $\Sigma$ and $\Upsilon$} \label{compissue}
To conduct statistical inference on $\theta_0$,
we provide two estimation methods for the Fisher information matrix $\Sigma$ or $\Upsilon$ in Theorems \ref{theorem2.1} -- \ref{theorem2.2}.

First, for $\Sigma$ in the stationary case, note that each entry of $\Sigma$ contains two components.
For example, the entry $\Sigma_{\phi_+ \phi_+ }:=\Sigma_{\sigma} \Sigma_\eta$, where
$\Sigma_\sigma=\frac{1}{4}E\big\{\frac{1}{\sigma_t^4(\theta_0)} \frac{\partial \sigma_t^2(\theta_0)}{\partial \phi_+ } \frac{\partial \sigma_t^2(\theta_0)}{\partial \phi_+ }\big\}$, and
$\Sigma_\eta=E\left[\big\{1+\frac{\partial \log f_{\alpha_0}(\eta)}{\partial x}\eta\big\}^2\right]$.
Given observations $\{y_0, y_1, ..., y_n\}$,
it is straightforward that
$\widehat{\Sigma}_\sigma:=\frac{1}{4n}\sum_{t=1}^{n} \frac{1}{\widetilde{\sigma}_t^4(\widehat{\theta}_{n})} \frac{\partial \widetilde{\sigma}_t^2(\widehat{\theta}_{n})}{\partial \phi_+ }
\frac{\partial \widetilde{\sigma}_t^2(\widehat{\theta}_{n})}{\partial \phi_+ }$.
As for $\Sigma_\eta$, we define two estimators, where
$\widehat{\Sigma}_\eta^{\mathrm{int}}$ is based on the integral expression of $\Sigma_\eta$, and $\widehat{\Sigma}_\eta^{\mathrm{res}}$ is based on the residuals $\widehat{\eta}_t=y_t/\widetilde{\sigma}_t(\widehat{\theta}_n)$:
\begin{flalign*}
	\widehat{\Sigma}_\eta^{\mathrm{int}}
	:=\int_{\mathbb{R}}\Big\{1+\frac{\partial \log f_{\widehat{\alpha}_n}(u)}
	{\partial x}u \Big\}^2f_{\widehat{\alpha}_n}(u)du, ~
	\widehat{\Sigma}_\eta^{\mathrm{res}}:=\frac{1}{n}\sum_{t=1}^n
	\Big\{1+\frac{\partial\log f_{\widehat{\alpha}_n}(\widehat{\eta}_t)}{\partial x}
	\widehat{\eta}_t\Big\}^2.
\end{flalign*}
By combining $\widehat{\Sigma}_\sigma$ with $\widehat{\Sigma}_\eta^{\mathrm{int}}$ and $\widehat{\Sigma}_\eta^{\mathrm{res}}$,
we obtain two estimators of
$\Sigma_{\psi \psi}$, say,  $\widehat{\Sigma}_{\psi \psi}^{\mathrm{int}}$
and $\widehat{\Sigma}_{\psi \psi}^{\mathrm{res}}$.
The remaining entries can be dealt with analogously, yielding two final estimators
$\widehat{\Sigma}^{\mathrm{int}}$ and $\widehat{\Sigma}^{\mathrm{res}}$.

As for $\Upsilon$ in the explosive case, the terms $E(d_t d_t^{\prime})$ and $E(d_t)$ depend only on the innovation $\eta$. 
For example, 
$E\{(d_{t}^{\psi})^2\}=\nu_{2} (1+\nu_{1} )/\{\psi_{0}^{2}
(1-\nu_{2}) (1-\nu_{1} )\}$, where 
$ \nu_{i}=E\left[\{\psi_{0}/a_{0}(\eta_{t})\}^{i}\right]$ for $i=1, 2$.
Thus, we similarly define
\begin{flalign*}
	\widehat{\nu}_{i}^{\mathrm{int}}&=\int_{\mathbb{R}}
	\Big\{\frac{\widehat{\psi}_{n}}{\widehat{\phi}_{n+}(u^{+})^2+\widehat{\phi}_{n-}(u^{-})^2+\widehat{\psi}_{n}}\Big\}^{i} f_{\widehat{\alpha}_n}(u)du,\\
	\widehat{\nu}_{i}^{\mathrm{res}}&=\frac{1}{n}\sum_{t=1}^n \Big\{\frac{\widehat{\psi}_{n}}{\widehat{\phi}_{n+}(\hat{\eta}_{t}^{+})^2
		+\widehat{\phi}_{n-}(\hat{\eta}_{t}^{-})^2+\widehat{\psi}_{n}}\Big\}^{i}, \quad i=1,2.
\end{flalign*}
The remaining part of $\Upsilon$ can be handled similarly.
Consequently, we obtain two corresponding estimators, denoted as
$\widehat{\Upsilon}^{\mathrm{int}}$ and $\widehat{\Upsilon}^{\mathrm{res}}$.
The following theorem shows that these estimators are all strongly consistent.

\begin{theorem} \label{theorem2.3}
	Suppose that Assumptions \ref{asm1}--\ref{asm2} hold.
	
	\noindent
	$(\mathrm{i})$. If $\gamma_{\alpha_0}<0$, then $\widehat{\Sigma}^{\mathrm{int}} \xrightarrow{a.s.} \Sigma$, and $\widehat{\Sigma}^{\mathrm{res}} \xrightarrow{a.s.} \Sigma$, as $n \rightarrow \infty$.
	
	\noindent
	$(\mathrm{ii})$. If $\gamma_{\alpha_0}>0$, then $\widehat{\Upsilon}^{\mathrm{int}} \xrightarrow{a.s.} \Upsilon$, and $\widehat{\Upsilon}^{\mathrm{res}} \xrightarrow{a.s.} \Upsilon$, as $n \rightarrow \infty$.
\end{theorem}

Numerical studies in Section \ref{simulation} show that both types of estimators
(i.e., $\widehat{\Sigma}^{\mathrm{int}}$ and $\widehat{\Sigma}^{\mathrm{res}}$,
$\widehat{\Upsilon}^{\mathrm{int}}$ and $\widehat{\Upsilon}^{\mathrm{res}}$) perform well, each with its own pros and cons.
Based on our findings, we provide a 
practical guidance:
for $\widehat{\alpha}_n\in(0, 1]$, the integration-based estimators  ( $\widehat{\Sigma}^{\mathrm{int}}$ or $\widehat{\Upsilon}^{\mathrm{int}}$) are recommended;
and for $\widehat{\alpha}_n \in (1, 2)$, the residual-based estimators
 ($\widehat{\Sigma}^{\mathrm{res}}$ or $\widehat{\Upsilon}^{\mathrm{res}}$) are preferable.

Regarding computation, since the density $f_\alpha(x)$ involves improper integrals of oscillating functions and lacks an explicit formula, numerical integration techniques are essential.
For all numerical calculations involving $f_\alpha(x)$, we recommend applying transformation techniques in \cite{Nolan1997}
and \cite{Matsui}.
The R code for the numerical calculation of stable densities is available at the GitHub link in \cite{Li:2023}.

\subsection{A universal estimator of the asymptotic variance} \label{universal}

In Theorems \ref{theorem2.1}--\ref{theorem2.2}, we establish the asymptotic variance of $\widehat{\vartheta}_n$ separately for the stationary and explosive cases.
However, in practice, the stationarity of the underlying process is typically unknown.
Importantly, valid inference for $\vartheta_0$ can still be conducted without imposing any stationarity condition.

Specifically, from Theorem \ref{theorem2.1}(ii), in the stationary case when $\gamma_{\alpha_0} < 0$, the asymptotic distribution of $\widehat{\vartheta}_{n}$ is
$\sqrt{n}(\widehat{\vartheta}_{n}-\vartheta_{0}) \xrightarrow{d}
\mathcal{N}(0, \Upsilon_{*}^{-1})$,
where
\begin{flalign*}
	\Upsilon_{*}=\Sigma_{\vartheta \vartheta}-\Sigma_{\vartheta \omega} \Sigma_{\omega \omega}^{-1}
	\Sigma_{\vartheta \omega}^{\prime}=
	\Bigg(
	\begin{array}{c c}
		\Sigma_{\widetilde{\vartheta} \widetilde{\vartheta}} & \Sigma'_{\widetilde{\vartheta} \alpha} \\
		\Sigma_{\widetilde{\vartheta} \alpha} & \Sigma_{\alpha\alpha}
	\end{array}
	\Bigg)-\Sigma_{\omega \omega}^{-1}\Bigg(
	\begin{array}{c}
		\Sigma_{\widetilde{\vartheta} \omega} \\
		\Sigma_{\alpha \omega}
	\end{array}
	\Bigg)
	\big(
	\Sigma_{\widetilde{\vartheta} \omega}' \quad \Sigma_{\alpha \omega} 
	\big).
\end{flalign*}
Define the residual-based estimator of $\Sigma_{\widetilde{\vartheta}\widetilde{\vartheta}}$ as follows:
\begin{flalign*}
		\begin{split}
			\widehat{\Sigma}_{\widetilde{\vartheta}\widetilde{\vartheta}}
			=\frac{1}{4}\Big\{\frac{1}{n}\sum_{t=1}^n\frac{1}{\widetilde{\sigma}_t^4(\widehat{\theta}_n)} \frac{\partial \widetilde{\sigma}_t^2(\widehat{\theta}_n)}{\partial \widetilde{\vartheta}} \frac{\partial \widetilde{\sigma}_t^2(\widehat{\theta}_n)}{\partial \widetilde{\vartheta}'}\Big\}
			\frac{1}{n}\sum_{t=1}^n\Big\{1+\frac{\partial \log f_{\widehat{\alpha}_n}(\widehat{\eta}_t)}{\partial x}\widehat{\eta}_t\Big\}^2, 
		\end{split}
\end{flalign*}
and other submatrices in $\Upsilon_{*}$ can be estimated accordingly.
Let $\widehat{\Upsilon}_{*}=\widehat{\Sigma}_{\vartheta \vartheta}-\widehat{\Sigma}_{\vartheta \omega} \widehat{\Sigma}_{\omega \omega}^{-1} \widehat{\Sigma}_{\vartheta \omega}^{\prime}$.
It can be shown that $\widehat{\Upsilon}_{*}$ is a consistent estimator of $\Upsilon_{*}$
in the stationary case.
The following theorem shows that $\widehat{\Upsilon}_{*}$ also consistently estimates the inverse asymptotic variance of $\widehat{\vartheta}_{n}$ in the explosive case.

\begin{theorem} \label{theorem2.4}
	Suppose that Assumptions \ref{asm1}--\ref{asm2} hold.
	
	\noindent
	$(\mathrm{i})$. If $\gamma_{\alpha_0}<0$, then $\widehat{\Upsilon}_{*} \xrightarrow{a.s.} \Upsilon_{*}$, as $n \rightarrow \infty$.
	
	\noindent
	$(\mathrm{ii})$. If $\gamma_{\alpha_0}>0$, then $\widehat{\Upsilon}_{*} \xrightarrow{a.s.} \Upsilon$, as $n \rightarrow \infty$.
\end{theorem}

In both cases, $\widehat{\Upsilon}_{*}^{-1}$ is a strongly consistent estimator of the asymptotic variance of $\widehat{\vartheta}_{n}$.
Thus, within this unified framework, we can conduct asymptotically valid inference for $\vartheta_0$ without requiring a prior stationarity test. The asymmetry test below is a direct application of this result.

\section{Testing}\label{test}
In this section, we discuss three types of hypothesis tests: stationarity testing,  asymmetry testing, and Kolmogorov-type test for diagnostic checking.

\subsection{Strict stationarity testing} \label{test.stationary}
To ensure valid inference for all estimators including $\omega_0$, we must study the stationarity of the underlying process. Consider the following two tests:
\begin{flalign} \label{4.1}
	&\text{(i)} \quad \text{Strict stationarity test: } \quad  H_{0}: \gamma_{\alpha_0}<0 \quad \text {vs} \quad H_{1}: \gamma_{\alpha_0} \geq 0; \\ 
	&\text{(ii)} \quad  \text{Explosivity test: } \quad H_{0}: \gamma_{\alpha_0} > 0 \quad \text {vs} \quad H_{1}: \gamma_{\alpha_0}\leq 0.   \label{4.2}
\end{flalign}
To estimate $\gamma_{\alpha_0}$, we use the residual-based estimator: $\widehat{\gamma}_{n}:=\frac{1}{n}\sum_{t=1}^n\log \big[\widehat{\phi}_{n+}$ $(\hat{\eta}_{t}^{+})^2+\widehat{\phi}_{n-}(\hat{\eta}_{t}^{-})^2+\widehat{\psi}_{n}\big]$.
Theorem \ref{theorem4.1} gives the asymptotic distribution of $\widehat{\gamma}_{n}$.
\begin{theorem}\label{theorem4.1}
	Let $u_{t}=\log a_{0}(\eta_{t})-\gamma_{\alpha_0}$, and $\sigma_{u}^{2}=E u_{t}^{2}< \infty$.
	Suppose that Assumptions \ref{asm1}--\ref{asm3} hold,
	it follows that
	\begin{flalign*}
		\sqrt{n}(\widehat{\gamma}_{n}-\gamma_{\alpha_0}) \xrightarrow{d}
		\mathcal{N}(0, \sigma_{\gamma}^{2}) \quad \text{as} \,\, n \rightarrow \infty,
	\end{flalign*}
	where
	$ \sigma_{\gamma}^{2}=
	\Biggl\{
	\begin{array}{ll}
		\sigma_{u}^{2}+\{a_1^{\prime} \Sigma^{-1} a_2 - 4(1-\nu_1)^2/c_1 \}, & \text{if} ~ \gamma_{\alpha_0}<0,\\
		\sigma_{u}^{2}, & \text{if} ~  \gamma_{\alpha_0} > 0,
	\end{array}
	\Biggr.
	$
    and the explicit forms of $a_1, a_2, \nu_1, c_1$ are given in the Appendix \ref{proof4.1}.
\end{theorem}
\begin{remark}
	Compared with Theorem 4.1 in \cite{FZ:2013}, 
	the asymptotic variance of $\widehat{\gamma}_{n}$ takes a more complicated form. 
	This difference arises from our distinct innovation assumptions, so the asymptotic behavior of $\widehat{\gamma}_{n}$ also depends on the estimation effect of stability parameter $\alpha$.
	The condition $\sigma_{u}^{2} = E u_{t}^{2}< \infty$ is ensured by the fact that $E(|\eta_t|^{\alpha_0/2})<\infty$.
\end{remark}
Denote
$\widehat{\sigma}_{u}^{2}=\frac{1}{n}\sum_{t=1}^n\big\{\log\big[\widehat{\phi}_{n+}(\hat{\eta}_{t}^{+})^2
+\widehat{\phi}_{n-}(\hat{\eta}_{t}^{-})^2+\widehat{\psi}_{n}\big]\big\}^2
-\big\{\frac{1}{n}\sum_{t=1}^n \\ \log\big[\widehat{\phi}_{n+}(\hat{\eta}_{t}^{+})^2
+\widehat{\phi}_{n-}(\hat{\eta}_{t}^{-})^2+\widehat{\psi}_{n}\big]\big\}^2$.
Under the assumptions of Theorem \ref{theorem4.1}, it can be shown
that $\widehat{\sigma}_{u}^{2}$ converges in probability to $\sigma_{u}^{2}$.
Construct the test statistic
$\mathrm{T}_{n}:=\sqrt{n} \widehat{\gamma}_{n} / \widehat{\sigma}_{u}
$.
Then, for the testing problem (\ref{4.1}) (resp., (\ref{4.2})),
the test defined by the stationary (resp., explosive) critical region
$
	\mathrm{C}^{\mathrm{ST}}=\{\mathrm{T}_{n}>\Phi^{-1}(1-\underline{\alpha})\}$ 
(resp., $\mathrm{C}^{\mathrm{EP}}=\{\mathrm{T}_{n}<\Phi^{-1}(\underline{\alpha})\}$),
has its asymptotic significance level bounded by $\underline{\alpha}\in(0, 1)$ and is consistent.

\subsection{Asymmetry testing}
It is particularly interesting to test for the existence of a leverage effect in financial asset returns.
Benefiting from the framework of our sAGARCH model (\ref{sgarch}), the asymmetry testing problem takes a simple form of
\begin{flalign} \label{4.3}
	H_{0}: \phi_{0+}=\phi_{0-} \quad \text{vs} \quad H_{1}: \phi_{0+} \neq \phi_{0-}.
\end{flalign}
Consider the following test statistic for symmetry:
\begin{flalign*}
	\mathrm{T}_{n}^{\mathrm{S}}=\frac{\sqrt{n}(\widehat{\phi}_{n+}-\widehat{\phi}_{n-})}{\widehat{\sigma}_{s}}, \quad\mbox{with}\,\,\,\, \widehat{\sigma}_{s}=\sqrt{\mathbf{e}^{\prime}
		\widehat{\Upsilon}_{*}^{-1} \mathbf{e}}
	\quad \mbox{and} \quad \mathbf{e}=(1,-1,0,0)'.
\end{flalign*}
Note that this symmetry test does not require any stationarity assumption, as it leverages the universal estimator $\widehat{\Upsilon}_{*}$ of the asymptotic variance of $\widehat{\vartheta}_n$. 
The following corollary holds directly from Theorem \ref{theorem2.4}.
\begin{corollary} \label{cor1}
	Suppose that Assumptions \ref{asm1}--\ref{asm3} hold. For the symmetry test (\ref{4.3}), the test defined by the critical region
	$\mathrm{C}^{\mathrm{S}}=\left\{|\mathrm{T}_{n}^{\mathrm{S}}|>\Phi^{-1}(1-\underline{\alpha} / 2)\right\}$
	has the asymptotic significance level $\underline{\alpha}\in(0, 1)$ and is consistent.
\end{corollary}

\subsection{Diagnostic checking}
Diagnostic checking is crucial in time series modeling, while the standard portmanteau test based on autocorrelations of residuals or squared residuals is unsuitable for assessing the adequacy of model (\ref{sgarch}), where the innovation is assumed to follow stable distribution. 
Here, we propose a Kolmogorov-type test for diagnostic checking. Consider the following null hypothesis:
\begin{flalign}\label{Bai_test}
	H_0: \eta_{t}\sim f_{\alpha_*}(x) \quad \mbox{ (i.e., }\alpha_{0}=\alpha_{*}\mbox{)},
\end{flalign}
where $\alpha_*\in(0, 2)$ is a fixed constant. In practice, $\alpha_{*}$ can be chosen to be $\widehat{\alpha}_{n}$ or an approximate value of
$\widehat{\alpha}_{n}$.

Let $F_\alpha(x)$ be the CDF of $\mathbf{S}(\alpha,0,1,0)$, $U_t=F_{\alpha_*}(\eta_t)$, and $V_n(r)=\frac{1}{\sqrt{n}}\sum_{t=1}^n [\mathbbm{1} (U_t\leq r )-r ]$ for $r\in[0, 1]$.
By the Donsker Theorem and continuous mapping theorem, under $H_{0}$, $\{U_t\}$ are i.i.d. $U[0, 1]$ r.v., and
\begin{flalign}\label{limiting_bai}
	\sup_{0\leq r\leq 1}|V_n(r)| \xrightarrow{d} 
	\sup_{0\leq r\leq 1} |\mathbb{B}(r)-r\mathbb{B}(1)|\quad \mbox{as $n\to\infty$},
\end{flalign}
where $\mathbb{B}(\cdot)$ is a standard Brownian motion.
Since $\eta_{t}$ is not observable in practice, we replace it with the residual $\widetilde{\eta}_{t}$,
where $\widetilde{\eta}_t=y_t/\sigma_t(\widetilde{\theta}_n)$,
$t=1,...,n$, and $\widetilde{\theta}_{n}=\big(\widehat{\omega}_n, \widehat{\phi}_{n+},
\widehat{\phi}_{n-}, \widehat{\psi}_n\big)'$ is
the restricted MLE of $\widetilde{\theta}_{0}$ under $H_0$.
Accordingly, we replace $U_t$ and $V_n(r)$ by $\widehat{U}_t$ and
$\widehat{V}_n(r)$ respectively, where
$\widehat{U}_t=F_{\alpha_*}(\widetilde{\eta}_t)$ and 
$\widehat{V}_n(r)=\frac{1}{\sqrt{n}}\sum_{t=1}^n\big[\mathbbm{1} \big(\widehat{U}_t\leq
r\big)-r\big]$, for $r\in[0, 1]$.

Unfortunately as opposed to (\ref{limiting_bai}),
$\sup _{0 \leq r \leq 1}|\widehat{V}_{n}(r)|$ is no longer asymptotically distribution-free,
since it involves the effect of parameter estimation.
To obtain an asymptotically distribution-free test statistic, we adopt the martingale transformation of \cite{Khmaladze} inspired by \cite{bai}. Specifically, let
$ g(r)=(g_{1}(r), g_{2}(r))^{\prime}=\big(r, f_{\alpha_{*}}(F_{\alpha_{*}}^{-1}(r)) F_{\alpha_{*}}^{-1}(r)\big)^{\prime} $
with its derivative $\dot{g}(r)=(1, \dot{g}_{2}(r))^{\prime}$, where
$\dot{g}_{2}(r)=1+ \frac{\dot{f}_{\alpha_*}(F^{-1}_{\alpha_*}(r))}{f_{\alpha_*}(F^{-1}_{\alpha_*}(r))} F_{\alpha_{*}}^{-1}(r)$ 
and $\dot{f}_{\alpha_*}(x)=\partial f_{\alpha_*}(x)/\partial x$.
Define Khmaladze's transformation
$\widehat{W}_{n}(r)=\widehat{V}_n(r)-
	\int_0^r (\dot{g}(s)'C^{-1}(s)\int_s^1\dot{g}(u)d\widehat{V}_n(u) )ds$,
where $C(r)=\int_{r}^{1} \dot{g}(s) \dot{g}(s)^{\prime} ds$. Then, the Kolmogorov-type test statistic is defined as
$\mathrm{T}_{n}^{\mathrm{D}}=\sup _{0 \leq r \leq 1}|\widehat{W}_{n}(r)|$.

\begin{theorem}\label{theorem4.2}
	Suppose model (\ref{sgarch}) is well specified with $\alpha_{0}=\alpha_{*}$ and 
	the Assumptions \ref{asm1}--\ref{asm3} hold.
	Then for both $\gamma_{\alpha_*}<0$ and $\gamma_{\alpha_*}>0$ cases,
	\begin{flalign*}
		\mathrm{T}_{n}^{\mathrm{D}} \xrightarrow{d} \sup _{0 \leq r \leq 1}|\mathbb{B}(r)| \quad \text { as } n \rightarrow \infty.
	\end{flalign*}
\end{theorem}
Thus, for the testing problem (\ref{Bai_test}), the critical region is defined as
$\mathrm{C}^{\mathrm{D}}=\big\{\mathrm{T}_{n}^{\mathrm{D}}>\mathrm{cv}_{\underline{\alpha}}^{\mathrm{D}}\big\}$
at the asymptotic significance level $\underline{\alpha}\in(0, 1)$, where $\mathrm{cv}_{\underline{\alpha}}^{\mathrm{D}}$ is the $(1-\underline{\alpha})$ quantile of the limiting distribution of $\mathrm{T}_{n}^{\mathrm{D}}$.
The critical values at the significance levels of 10\%,
5\%, and 1\% are 1.9571, 2.2340, and 2.8080, respectively, via simulation.
Numerical simulations in Section \ref{simulation}
show that $\mathrm{T}_{n}^{\mathrm{D}}$ has satisfactory power in all cases, even for small samples.

\begin{remark}
	In application, $\mathrm{T}_{n}^{\mathrm{D}}$ can be approximately computed with
	\begin{flalign*}
		\max_{1\leq j\leq n}\sqrt{n} \bigg|\frac{j}{n}-\frac{1}{n}\sum_{k=1}^j\dot{g}(v_k)'C_k^{-1}D_k(v_k-v_{k-1}) \bigg|,
	\end{flalign*}
	where $D_k=\sum_{i=k}^n\dot{g}(v_i)$, $C_k=\sum_{i=k}^n\dot{g}(v_i)\dot{g}(v_i)'(v_{i+1}-v_{i})$,
	and $v_1,...,v_n$ are ordered values of $\widehat{U}_1,...,\widehat{U}_n$ with the convention $v_{0}=0$ and $v_{n+1}=1$.
\end{remark}

\section{Simulation Studies}\label{simulation}
\subsection{Performance of the MLE}
To assess the finite-sample performance of the MLE of $\theta_0$ in the sAGARCH model (\ref{sgarch}),
we choose $n=200, 500, 1000, 2000$ with 1000 replications.
\revise{First, we consider three stationary cases corresponding to  $\alpha_0=1.5, 1.0$, and $0.5$.
To streamline the presentation, Table \ref{table1} reports only the case $\alpha_0=1.5$, which is close to the empirical tail index in Section \ref{example}. 
The complete results for all three values of $\alpha_0$ are provided in the Appendix \ref{appA1}.}

\begin{table}[!t]
	\revise{
	\caption{\label{table1} Simulation results for the MLE $\widehat{\theta}_n$ for sAGARCH(1,1) under the stationary case with $\alpha_0=1.5$.}}
	\centering
	\resizebox{0.85\linewidth}{!}{
			\begin{tabular}{c|l|ccccc}
				\hline
				\hline
				\multicolumn{1}{c|}{\multirow{2}{*}{$n$}}
				&
				& \multicolumn{1}{c}{$\omega_0=0.2$}
				& \multicolumn{1}{c}{$\phi_{0+}=0.1$}
				& \multicolumn{1}{c}{$\phi_{0-}=0.2$}
				& \multicolumn{1}{c}{$\psi_0=0.5$}
				& \multicolumn{1}{c}{$\alpha_0=1.5$} \\
				\cline{3-7}
				\multicolumn{1}{c|}{}
				&
				& \multicolumn{1}{c}{$\widehat{\omega}_n$}
				& \multicolumn{1}{c}{$\widehat{\phi}_{n+}$}
				& \multicolumn{1}{c}{$\widehat{\phi}_{n-}$}
				& \multicolumn{1}{c}{$\widehat{\psi}_n$}
				& \multicolumn{1}{c}{$\widehat{\alpha}_n$} \\
				\hline
				\hline
				\multicolumn{1}{c|}{\multirow{5}{*}{$200$}}
				& Bias
				& 0.0360 & 0.0038 & 0.0008 & -0.0007 & 0.0253 \\
				\multicolumn{1}{c|}{}
				& ESD
				& 0.1213 & 0.0610 & 0.0883 & 0.1086 & 0.1131 \\
				\multicolumn{1}{c|}{}
				& ASD
				& 0.0890 & 0.0504 & 0.0813 & 0.0878 & 0.1085 \\
				\multicolumn{1}{c|}{}
				& $\widehat{\mathrm{ASD}}^{\mathrm{int}}$
				& 0.1095 & 0.0519 & 0.0818 & 0.0948 & 0.1070 \\
				\multicolumn{1}{c|}{}
				& $\widehat{\mathrm{ASD}}^{\mathrm{res}}$
				& 0.1099 & 0.0521 & 0.0821 & 0.0951 & 0.1076 \\
				\hline
				\multicolumn{1}{c|}{\multirow{5}{*}{$500$}}  
				& Bias                                    
				& 0.0181 & 0.0001 & -0.0026 & 0.0016 & 0.0113 \\
				\multicolumn{1}{c|}{} 
				& ESD                                     
				& 0.0653 & 0.0330 & 0.0518 & 0.0585 & 0.0712 \\
				\multicolumn{1}{c|}{} 
				& ASD                                     
				& 0.0563 & 0.0319 & 0.0514 & 0.0555 & 0.0687 \\
				\multicolumn{1}{c|}{} 
				& $\widehat{\mathrm{ASD}}^{\mathrm{int}}$ 
				& 0.0627 & 0.0320 & 0.0511 & 0.0576 & 0.0683 \\
				\multicolumn{1}{c|}{} 
				& $\widehat{\mathrm{ASD}}^{\mathrm{res}}$ 
				& 0.0628 & 0.0320 & 0.0512 & 0.0576 & 0.0684 \\
				\hline
				\multicolumn{1}{c|}{\multirow{5}{*}{$1000$}} 
				& Bias                                    
				& 0.0080 & 0.0008 & 0.0005 & -0.0002 & 0.0052 \\
				\multicolumn{1}{c|}{} 
				& ESD                                     
				& 0.0435 & 0.0239 & 0.0361 & 0.0409 & 0.0474 \\
				\multicolumn{1}{c|}{} 
				& ASD                                     
				& 0.0398 & 0.0225 & 0.0364 & 0.0392 & 0.0485 \\
				\multicolumn{1}{c|}{} 
				& $\widehat{\mathrm{ASD}}^{\mathrm{int}}$ 
				& 0.0422 & 0.0226 & 0.0366 & 0.0397 & 0.0485 \\
				\multicolumn{1}{c|}{} 
				& $\widehat{\mathrm{ASD}}^{\mathrm{res}}$ 
				& 0.0423 & 0.0227 & 0.0367 & 0.0397 & 0.0485 \\
				\hline
				\multicolumn{1}{c|}{\multirow{5}{*}{$2000$}} 
				& Bias                                    
				& 0.0014 & -0.0002 & -0.0005 & 0.0005 & 0.0025 \\
				\multicolumn{1}{c|}{} 
				& ESD                                     
				& 0.0284 & 0.0164 & 0.0253 & 0.0276 & 0.0328 \\
				\multicolumn{1}{c|}{} 
				& ASD                                     
				& 0.0282 & 0.0159 & 0.0257 & 0.0278 & 0.0343 \\
				\multicolumn{1}{c|}{} 
				& $\widehat{\mathrm{ASD}}^{\mathrm{int}}$ 
				& 0.0288 & 0.0159 & 0.0256 & 0.0279 & 0.0343 \\
				\multicolumn{1}{c|}{} 
				& $\widehat{\mathrm{ASD}}^{\mathrm{res}}$ 
				& 0.0288 & 0.0159 & 0.0256 & 0.0279 & 0.0343 \\
				\hline 
				\hline
	\end{tabular}}
\end{table}

\revise{
Table \ref{table1} reports the empirical biases (Bias), empirical standard deviations (ESD),
asymptotic standard deviations (ASD) of the MLE $\widehat{\theta}_n$, together with two estimators
$\widehat{\mathrm{ASD}}^{\mathrm{int}}$ and $\widehat{\mathrm{ASD}}^{\mathrm{res}}$ of ASDs
from $\widehat{\Sigma}^{\mathrm{int}}$ and $\widehat{\Sigma}^{\mathrm{res}}$ in Section \ref{compissue}.
The ASDs are calculated from Theorem \ref{theorem2.1}~(ii).
Table \ref{table1} shows that the biases are small and decrease with $n$, while the ESDs, ASDs, and estimated ASDs become close as $n$ increases.
The two estimators of the ASD perform similarly well. 
For heavier-tailed cases, the results are similar for most parameters, but $\widehat{\omega}_n$ may be less stable when $\alpha_0\leq 1$. 
According to our model, it is likely related to the extreme values of the observations.
Since the intercept $\omega$ can be viewed as a \textit{scale parameter} for $y_t$, the estimator of $\omega$ is likely influenced by the scale of $y_t$ and may be overestimated in some cases;
see Appendix \ref{appA1} for more discussions.}

Fig. \ref{fig2} plots the histograms of
$\sqrt{n}(\widehat{\theta}_n-\theta_0)$ with $n = 1000$ and $\theta_0=(0.2,0.1,0.2,0.5, 1.5)'$. 
\begin{figure}[!htbp]
	\begin{center}
		\includegraphics[width = 13cm]{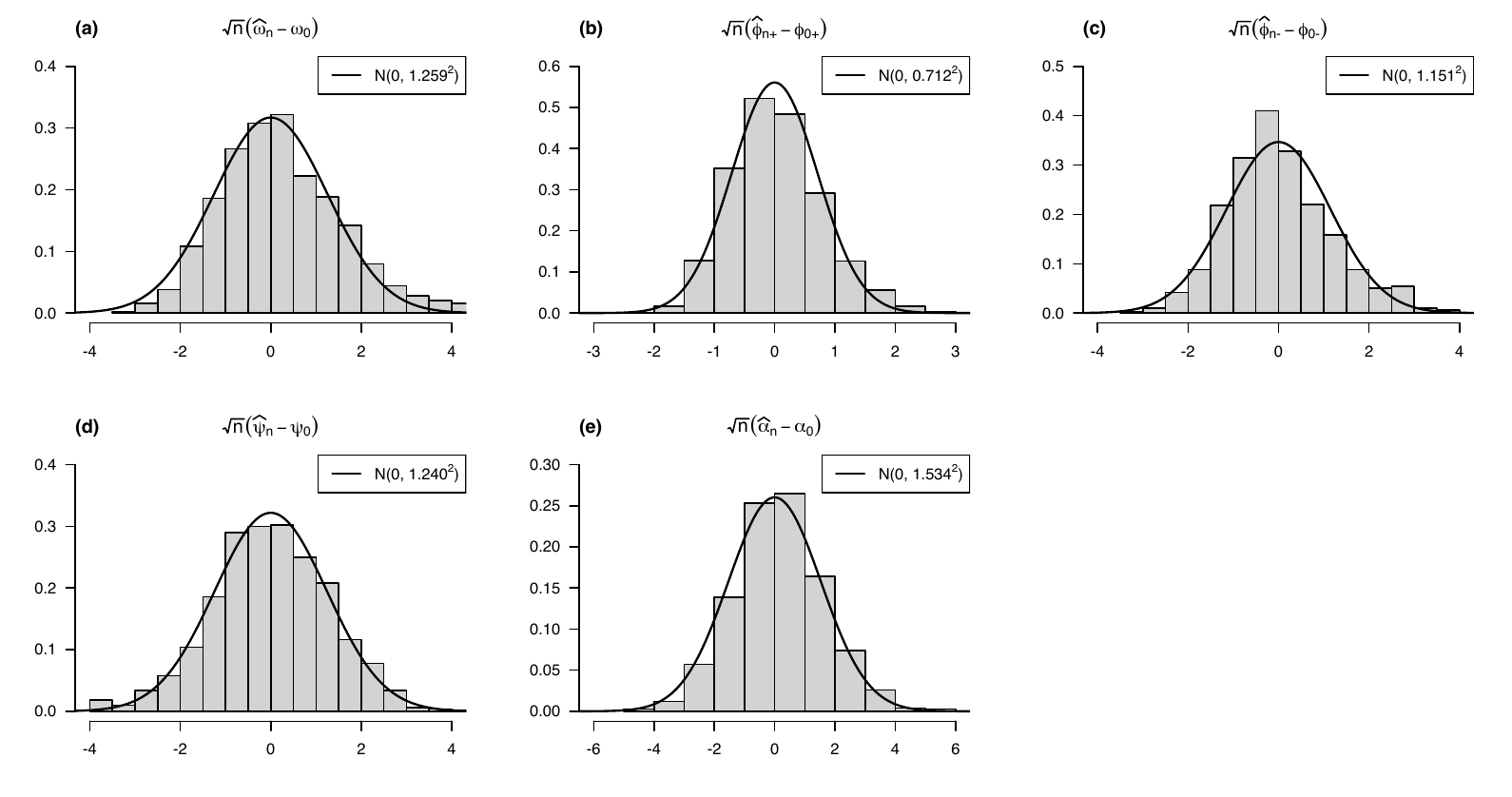}
		\caption{The histograms of $\sqrt{n}(\widehat{\theta}_n-\theta_0)$ with $n = 1000$ and $\theta_0=(0.2,0.1,0.2,0.5, 1.5)'$. }\label{fig2}
	\end{center}
\end{figure}
The results demonstrate that $\widehat{\theta}_n$ performs well overall
and conforms to the asymptotic normal distribution.
Further results on the finite-sample performance of MLE in the explosive cases and the estimation of the Lyapunov exponents are also presented in the Appendix \ref{appA2}.

\subsection{Performance of the test statistics}

Next, we examine the performance of diagnostic checking.
The results for the stationarity and asymmetry tests are presented in Appendix \ref{appA3}.

For the stationary cases, consider two scenarios: (I) sAGARCH(1,1) model
with $\theta_0=(0.2,0.1, 0.2, 0.5, 1.5)'$.
The null hypothesis is $H_0^{D}: \eta_t\sim f_{1.5}$, motivated by the empirical example in Section \ref{example}, and the alternative $H_1: \eta_t\sim f_{\alpha_*}$ with $\alpha_*\neq 1.5$.
(II) sAGARCH(1,1) model
with $\theta_0=(0.1,0.1, 0.2, 0.3, 1)'$. The null $H_0^{D}: \eta_t\sim f_{1}$
(i.e., standard Cauchy), with alternative $\eta_t\sim$ Student's $t_\nu$-distribution with $\nu\in[0.5, 5]$.
\begin{figure}[!t]
	\begin{center}
		\includegraphics[width = 12cm]{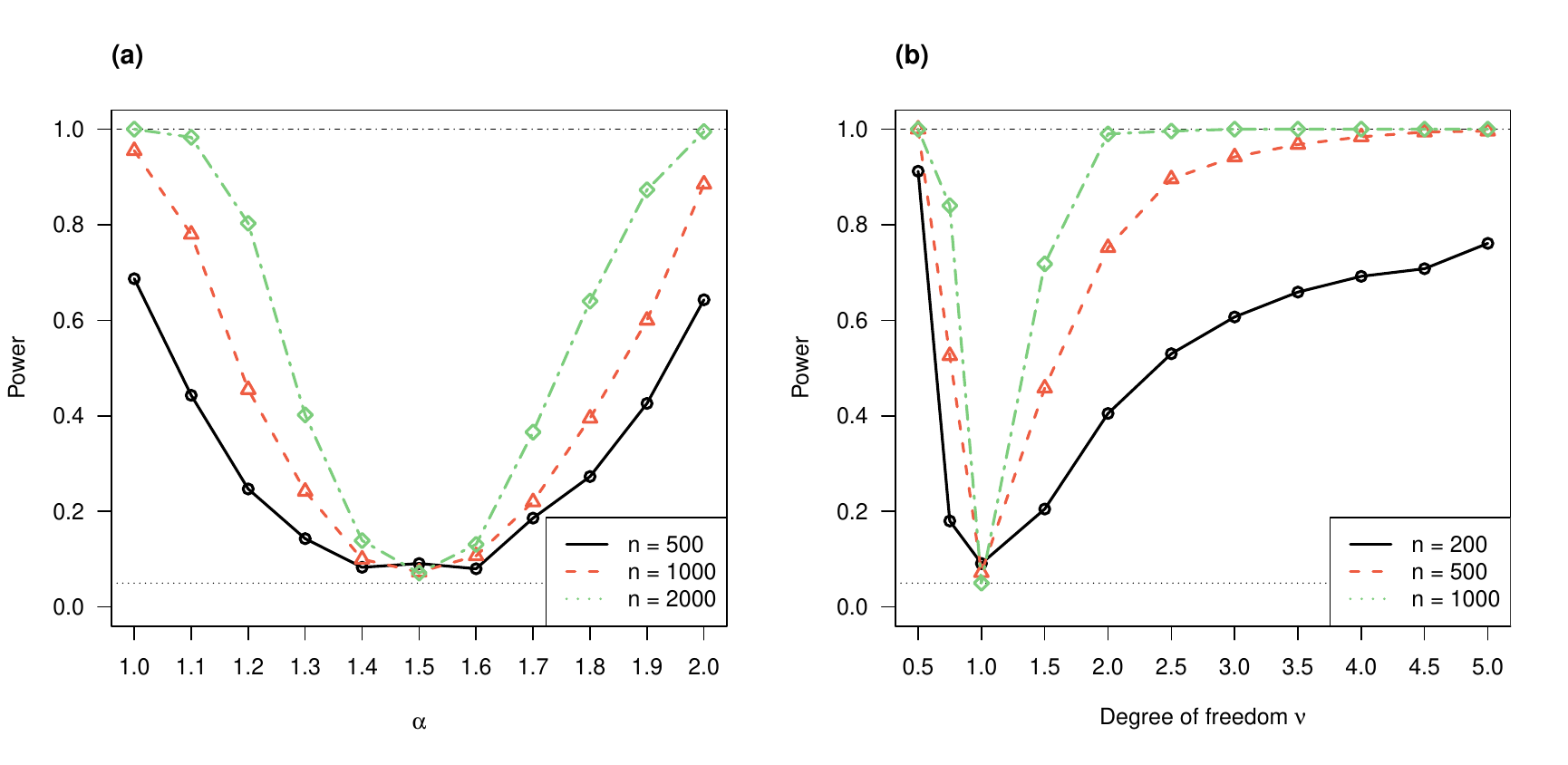}
		\caption{The size and power of $\mathrm{T}_{n}^{\mathrm{D}}$ for stationary cases. 
			(a) $H_1: \eta_t \sim f_\alpha(x)$ with $\alpha\in[1, 2]$. The size corresponds to $\alpha=1.5$.
			(b) $H_1: \eta_t\sim$ Student's $t_\nu$-distribution with $\nu\in[0.5, 5]$. The size corresponds to $\nu=1$  (i.e., $\eta_t\sim f_{1}$). The horizontal dotted line at the bottom denotes the 5\% significance level.}\label{fig4}
	\end{center}
\end{figure}
Fig. \ref{fig4} reports the size and power of $\mathrm{T}_{n}^{\mathrm{D}}$ at the significance level 5\%. Here $n$ is 500, 1000, 2000 for (I), and 200, 500,1000 for (II), each with 1000 replications.
The size is generally close to the nominal level, except for $n=500$ in (I), and the power increases as the stable exponent $\alpha$ or the degree of freedom $\nu$ deviates from the null. The size and power improve with increasing $n$.

We further examine the performance of $\mathrm{T}_{n}^{\mathrm{D}}$ under explosive cases. 
For sAGARCH(1,1) model with $\theta_0=(0.1,0.1, 0.2,$ $0.5, 1)'$, consider two alternatives to the null $H_0^D:\eta_t\sim f_1$:
(I) $\eta_t\sim f_{\alpha_*}$ with $\alpha_*\neq 1$, and
(II) $\eta_t\sim t_\nu$ with $\nu\in[0.5,5]$.
\begin{figure}[!t]
	\begin{center}
		\includegraphics[width = 12cm]{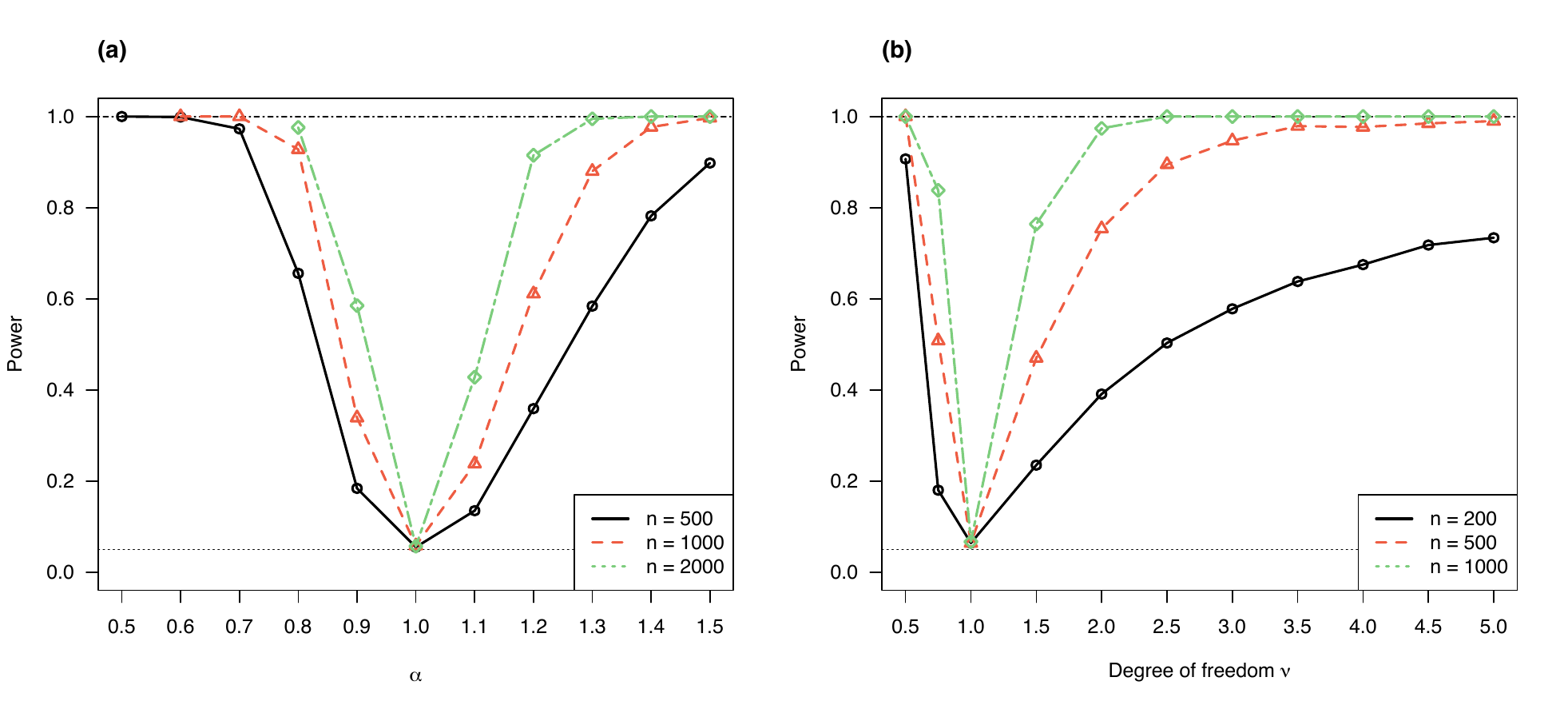}
		\caption{The size and power of $\mathrm{T}_{n}^{\mathrm{D}}$ for explosive cases. 
			(a) $H_1: \eta_t\sim f_\alpha(x)$, with $\alpha\in[0.5, 1.5]$;
			(b) $H_1: \eta_t\sim$ $t_\nu$-distribution, with $\nu\in[0.5, 5]$.
			}\label{fig5}
	\end{center}
\end{figure}
Fig. \ref{fig5} reports the corresponding results for explosive cases, again based on 1000 replications and the same sample sizes.
The test shows satisfactory size and increasing power as the innovation distribution moves away from the null specification.
In particular, (II) demonstrates that $\mathrm{T}_{n}^{\mathrm{D}}$ can effectively distinguish Student's $t$ innovations from misspecified stable innovations, confirming its usefulness for finite-sample diagnostic checking.

\section{Empirical Examples}\label{example}

\begin{figure}[!htbp]
	\begin{center}
		\includegraphics[width = 10cm]{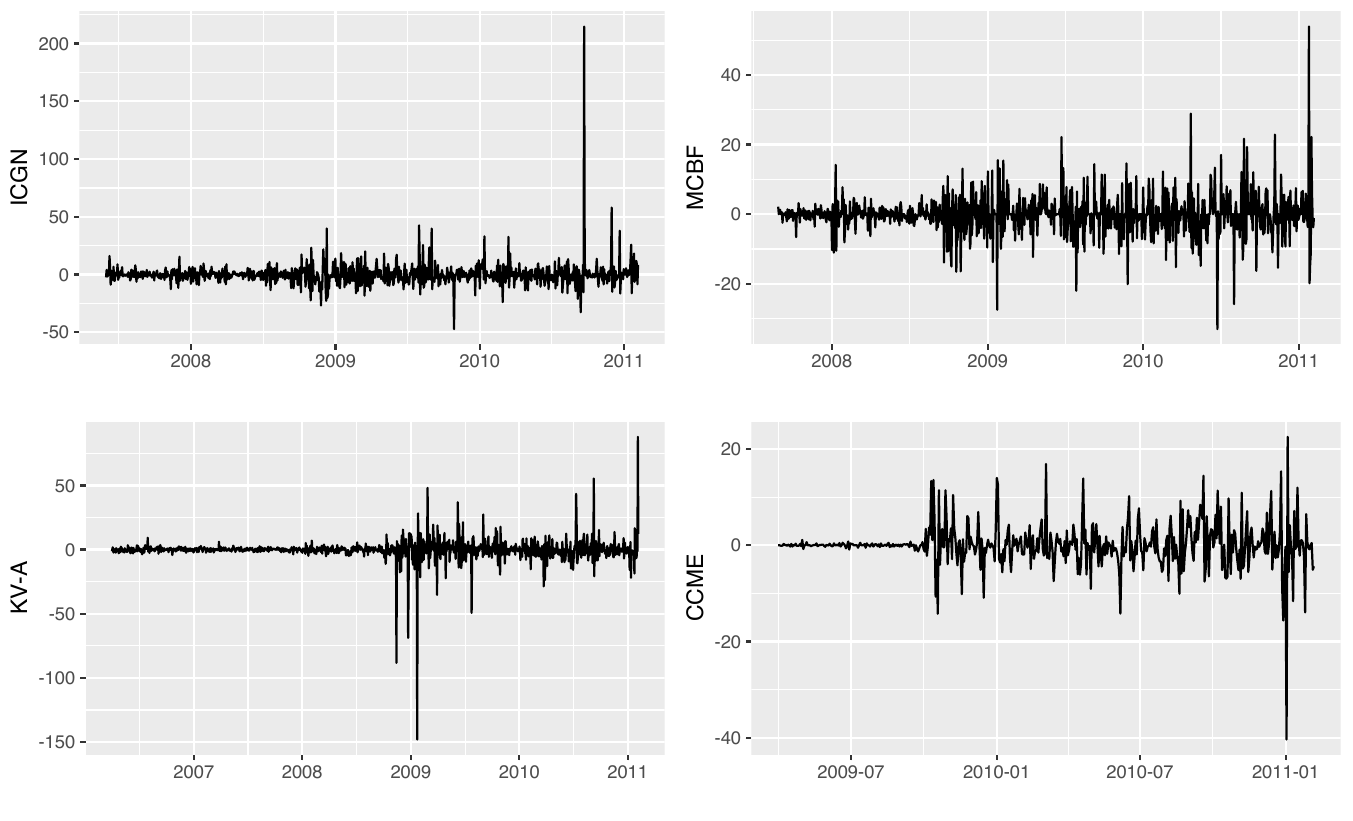}
		\caption{The graphs of four individual stock return series (\%).
			}\label{fig6}
	\end{center}
\end{figure}

In this section, we analyze four individual stock return series, which are the same as those in \cite{FZ:2012,FZ:2013} for ease of comparison.
They are the daily series of Icagen (NasdaqGM: ICGN, May 31, 2007--Feb. 7, 2011),
Monarch Community Bancorp (NasdaqCM: MCBF, Aug. 28, 2007 -- Feb. 7, 2011),
KV Pharmaceutical (NYSE: KV-A, Mar. 31, 2006 -- Feb. 7, 2011),
and China MediaExpress (NasdaqGS: CCME, Mar. 31, 2009 -- Feb. 7, 2011).
Fig. \ref{fig6} displays the four stock return series.

\revise{
We fit an sAGARCH(1,1) model  to each series and report the results in Table \ref{table5},
including parameter estimates, the explosivity test $\mathrm{T}_{n}$, asymmetry test $\mathrm{T}_{n}^{\mathrm{S}}$, and diagnostic statistics $\mathrm{T}_n^\mathrm{D}$, all with corresponding $p$-values.
The estimated ASDs in parentheses are computed using the universal estimator $\widehat{\Upsilon}_{*}$ in Section \ref{universal}.
We further conduct out-of-sample VaR/ES backtesting.
Define
$\widehat{\mathrm{VaR}}_{t+1}(\tau) = -
\sigma_{t+1}(\widehat{\theta}_n) q_{\widehat{\alpha}_n}(\tau)$,  
$\widehat{\mathrm{ES}}_{t+1}(\tau) = -
\sigma_{t+1}(\widehat{\theta}_n) E [\eta | \eta \leq q_{\widehat{\alpha}_n}(\tau)]$, where $q_{\alpha}(\tau)$ is the $\tau$-quantile of $\eta \sim f_\alpha(\cdot)$.
We use the first 80\% of the sample for estimation, the remaining 20\% for one-step-ahead VaR/ES forecasting, and set $\tau=10\%$.
We compute three backtesting statistics:
Kupiec's unconditional coverage test applied to the exceedance indicators
$I_t = \mathbbm{1} \{-y_t > \widehat{\mathrm{VaR}}_{t+1}(\tau)\}$, 
Christoffersen's independence test to examine clustering in $\{I_t\}$, and 
McNeil-Frey ES test of whether the excess conditional shortfall is i.i.d. with zero mean.
Rejection of any of the three null hypotheses indicates that the risk model is misspecified.
For comparison, we also report the QMLE results for the asymmetric GARCH(1,1) model in \cite{FZ:2013}, and the MLE results for two benchmark models that incorporate asymmetry in the innovation: 
a skew-$t$ GARCH(1,1) model and a skew-stable GARCH(1,1) model that allows skew-stable innovation ($\beta \neq 0$).
To save space, we report the AIC, computational time (s), and the minimum p-value among three VaR/ES backtests (``VaR/ES'').
For AGARCH(1,1), $p$-values of nonstationarity test (``$p$-value-$\mathrm{T}_{n}$'') and asymmetry test (``$p$-value-$\mathrm{T}_{n}^{\mathrm{S}}$ '') are also reported.
Bold entries highlight notable differences in testing results or models with better performance.}

\begin{table}[!t]
	\revise{\caption{\label{table5} 
	Empirical comparison of sAGARCH(1,1) and benchmark GARCH-type models.
	Notable differences in test results and the best-fitting models are highlighted in bold.}}
	\vspace{0.2cm}
	\resizebox{\linewidth}{!}{\begin{tabular}{cccccc} \hline   \hline 
		Model &   & ICGN         & MCBF         & KV-A                 & CCME         \\    \hline   \hline
		\multirow{11}{*}{\tabincell{c}{sAGARCH\\(1,1)}}  & $n$                      & 928          & 868          & 1221              & 488          \\
		& $\widehat{\phi}_{n+}$              & 0.098 (0.026) & 0.022 (0.008) & 0.033 (0.008) & 0.084 (0.022) \\
		& $\widehat{\phi}_{n-}$           & 0.164 (0.035) & 0.029 (0.008) & 0.039 (0.008)  & 0.090 (0.027) \\
		& $\widehat{\psi}_n$               & 0.419 (0.056) & 0.884 (0.022) & 0.835 (0.018) & 0.766 (0.034) \\
		& $\widehat{\alpha}_n$            & 1.556 (0.057) & 1.369 (0.065) & 1.587 (0.045) & 1.527 (0.084) \\
		& $p$-value-$\mathrm{T}_{n}$             & 0.000        & \textbf{0.056}        & \textbf{0.000}              & 0.285        \\
		& $p$-value-$\mathrm{T}_{n}^{\mathrm{S}}$                 & \textbf{0.108 }       & 0.532        & 0.527             & 0.849        \\
		& 	$p$-value-$\mathrm{T}_{n}^{\mathrm{D}}$                  &   0.014      &     0.000     &     0.588        &   0.045 \\
		& AIC                    & \textbf{6011.4}       & 4987.2       & \textbf{6278.1 }         & \textbf{2185.4   }    \\    
		& comp-time  &   56.3  &   39.8  &      83.4      &   36.7    \\  
		& VaR/ES  &   0.106   &   0.149   &   0.122      &    0.384    \\ \hline
		\multirow{5}{*}{\tabincell{c}{AGARCH\\(1,1) with \\ QMLE}} &  $p$-value-$\mathrm{T}_{n}$  & 0.008       & \textbf{0.515 }      & \textbf{0.708}         & 0.611     \\
		& $p$-value-$\mathrm{T}_{n}^{\mathrm{S}}$           & \textbf{0.037}        & 0.850        & 0.052           & 0.503        \\
		& AIC            & 6430.5       & 5275.5       & 7395.7          & 2272.1      \\  
		& comp-time  &   0.2   &   0.3  &    0.03       &   0.2    \\ 
		& VaR/ES  &  \textbf{0.041}   &   \textbf{0.009}    &   \textbf{ 0.001}      &    \textbf{0.028}    \\ \hline
		\multirow{3}{*}{\tabincell{c}{skew-$t$ \\ GARCH(1,1)}} 
		& AIC            & 6056.0       &  \textbf{4966.3}   &     6431.2      &   2248.2    \\  
		& comp-time  &   4.6   &  5.6    &     7.9      &   3.1    \\ 
		& VaR/ES  &   \textbf{0.012}  &    \textbf{0.098}   &    \textbf{0.040}      &   \textbf{0.020}     \\ \hline
		\multirow{3}{*}{\tabincell{c}{skew-stable \\ GARCH(1,1)}} 
		& AIC            &  \textbf{ 6013.1  }   &  4987.2   &    \textbf{ 6277.5}      &  \textbf{2183.1}     \\
		& comp-time  &   109.4   &   217.9   &    234.1      &   70.8    \\
		& VaR/ES  &   \textbf{0.012}   &   0.151   &   0.811       &    0.533    \\ 
		\hline  \hline
	\end{tabular}}
\end{table}

Table \ref{table5} reveals several notable findings:
\noindent
\revise{
(i) First, the sAGARCH(1,1) model provides a competitive fit. 
All estimated parameters are significant, and its AIC values are lower than those of the AGARCH(1,1) and skew-$t$ GARCH(1,1) models, except for MCBF. 
The skew-stable GARCH(1,1) model performs comparably in terms of AIC, but requires substantially more computational time due to the optimization of an additional skewness parameter and the complexity of stable densities.
It demonstrates that incorporating asymmetry in the volatility process $\sigma_t$ achieves similar fitting performance while reducing computational cost.}

\revise{
\noindent
(ii) Second, the testing results highlight the role of heavy-tailed modeling. 
For stationarity testing, the results for ICGN and CCME are consistent with those in \cite{FZ:2013},
whereas KV-A and MCBF lead to different conclusions under the stable-innovation framework. 
For ICGN, the null hypothesis of symmetry is not rejected at the 5\% level using our model, whereas the AGARCH(1,1) model in \cite{FZ:2013} detects asymmetry.
This indicates that apparent explosivity or leverage effects may be sensitive to the assumed innovation distribution and may partly reflect the influence of extreme observations.
In fact, after removing an obvious outlier around September 22, 2010 from ICGN, the series is tested to be symmetric even under the  AGARCH(1,1) model.}

\revise{
	\noindent
(iii) Third, the values of $\mathrm{T}_{n}^{\mathrm{D}}$ for the ICGN, KV-A, and CCME are below the critical values of $\sup_{0\leq r\leq 1}|\mathbb{B}(r)|$ at the significance levels of either 1\% or 5\%. In contrast, the values for the MCBF are relatively large.
This discrepancy is likely due to the apparent presence of change points in the MCBF time series, as seen in Fig. \ref{fig6}, which also contributes to the series' nonstationarity. Specifically, volatility increases sharply around mid-2008, possibly related to the financial crisis during that period. Testing for change points is thus an interesting topic for future research.}

\revise{
	\noindent
(iv) Fourth, the VaR/ES backtesting results support the adequacy of the proposed model for tail-risk evaluation. 
For sAGARCH(1,1), all reported VaR/ES $p$-values exceed $0.1$, suggesting no evidence of misspecification in the tail-risk forecasts. 
In contrast, the AGARCH(1,1) and skew-$t$ GARCH(1,1) models reject at least one VaR/ES backtest in most cases, and the skew-stable GARCH(1,1) performs worse than sAGARCH(1,1) for ICGN.}

\begin{figure}[!b]
	\begin{center}
		\includegraphics[width = 10cm]{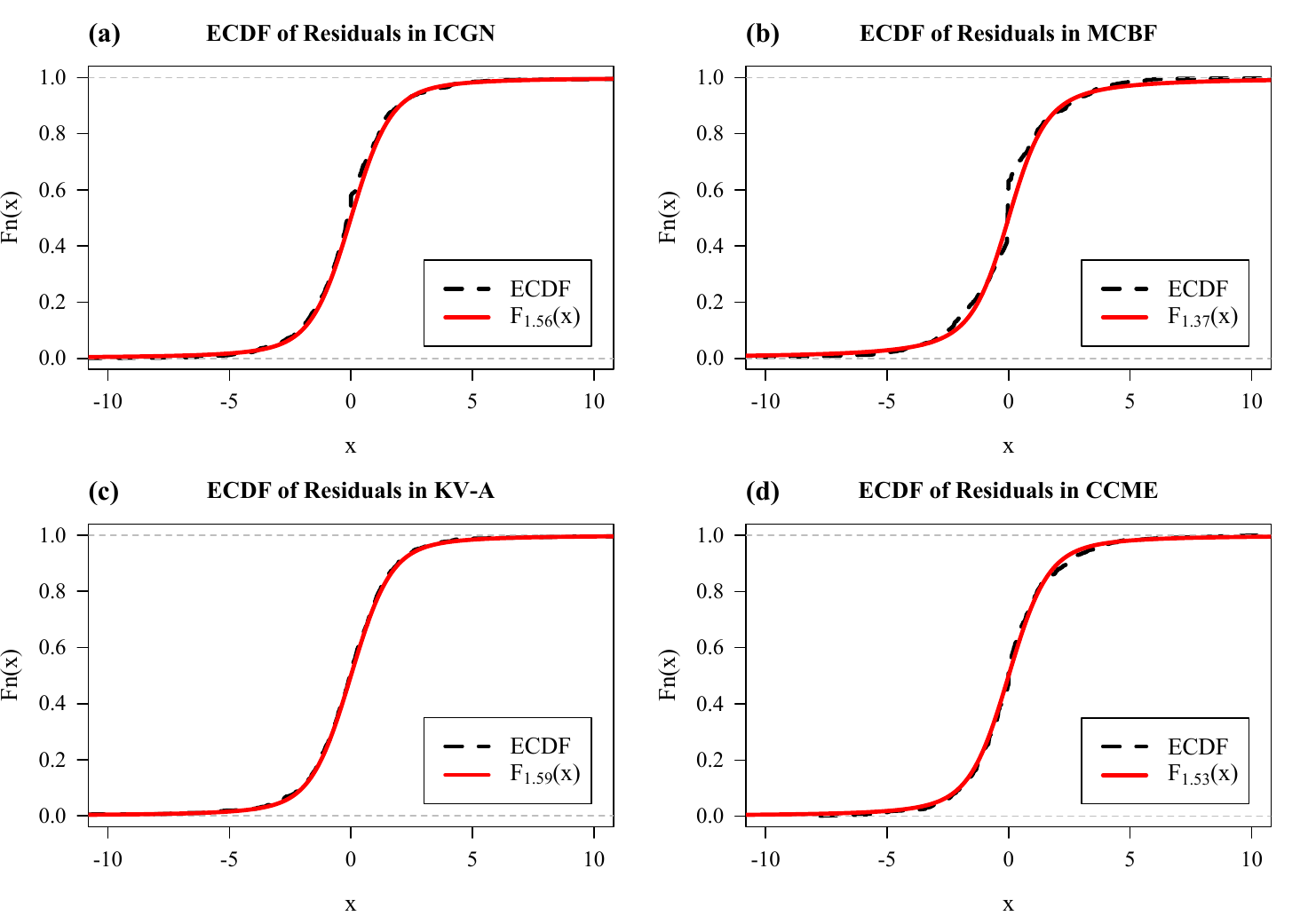}
		\caption{The ECDFs of residuals of four stock return series (\%).
			}\label{fig7}
	\end{center}
\end{figure}

We further examine the model-fitting performance of sAGARCH(1,1) by analyzing residuals in Fig. \ref{fig7}.
It compares the CDF of the stable distribution (with estimated $\widehat{\alpha}_n$) against the empirical CDF (ECDF) of the residual.
For ICGN, KV-A and CCME, the empirical CDF closely aligns with the true one, illustrating the goodness-of-fit of sAGARCH(1,1) model for these datasets. 
As for MCBF, there exists a certain difference between two CDFs, which 
suggests room for improvement--for instance, by incorporating change points into the model. Despite this, our sAGARCH(1,1) model still outperforms the original AGARCH(1,1) model.


We also evaluate the performance of model \eqref{sgarch} against the AGARCH model with $t$-innovations on several well-known portfolio returns in Appendix \ref{appA4}. 
The results illustrate the merits of our model in capturing aggregate market behavior and more effectively characterizing portfolio risk.

\section{Concluding Remarks}\label{conclusion}
In this paper, we propose a novel sAGARCH model that operates without moment conditions. 
The model effectively captures excess kurtosis, volatility clustering, and leverage effects of financial return series.
We develop a comprehensive statistical inference framework for the sAGARCH model, unifying both stationary and explosive cases. This addresses a critical gap in the literature on inference of heavy-tailed GARCH-type models.
Through extensive simulations, we uncover that the intercept estimator in heavy-tailed GARCH-type models can exhibit poor finite-sample performance. This finding highlights potential risks in financial return forecasting. 

To conclude, we outline several promising extensions.
First, generalizing the sAGARCH(1,1) model to higher-order specifications (sAGARCH($p, q$)) is natural. 
As addressed in \cite{chng2009}, such an extension would involve technical challenges in deriving the statistical inference theory.
Second, over-parametrization arises in estimating sAGARCH($p, q$) models.
When true parameters lie on the boundary of parameter space, inference will become nonstandard and more complicated.
Third, return series often exhibit structural breaks, making change point detection important for sAGARCH($p, q$) modeling. 
Finally, the asymptotic behavior of sAGARCH(1,1) model in the critical case ($\gamma_{\alpha_0} = 0$) remains an open question.

\section*{Supplementary Materials}

The Supplementary Material contains additional simulation results, empirical analyses on portfolio returns, and proofs of all theorems.
\par

\section*{Acknowledgements}

The authors are very grateful to the co-editor, the associate editor, and two anonymous referees for their constructive suggestions and comments, which led to significant improvement.
We would like to thank Professor Howell Tong for his insightful suggestions. 
Tao's research is supported by the NSFC (No.72503086) and the Natural Science Foundation of Guangdong Province (No.2025A1515010697).
Gong's research is supported by the NSFC (No.72501294) and the Innovation Research Foundation of NUDT (No.ZK25-61).
Li's research is supported by the NSFC (No.72471127).

\par


\bibhang=1.7pc
\bibsep=2pt
\fontsize{9}{14pt plus.8pt minus .6pt}\selectfont
\renewcommand\bibname{\large \bf References}
\expandafter\ifx\csname
natexlab\endcsname\relax\def\natexlab#1{#1}\fi
\expandafter\ifx\csname url\endcsname\relax
  \def\url#1{\texttt{#1}}\fi
\expandafter\ifx\csname urlprefix\endcsname\relax\def\urlprefix{URL}\fi


\begin{thebibliography}{4}
	
	\bibitem[\protect\citeauthoryear{Andrews et al.} {2009}]{acd}
	Andrews, B.,  Calder, M. and  Davis,  R.A. (2009). Maximum likelihood estimation for $\alpha$-stable autoregressive processes.
	\textit{Ann. Statist.} \textbf{37}(4),  1946--1982.
	
	
	\bibitem[\protect\citeauthoryear{Bai} {2003}]{bai}
	Bai, J. (2003). Testing parametric conditional distributions of dynamic models.
	\textit{Review of Economics and Statistics} \textbf{85}(3), 531--549
	
	
	\bibitem[\protect\citeauthoryear{Bollerslev}{1986}]{Bollerslev:1986}
	Bollerslev, T. (1986). Generalized autoregressive conditional heteroskedasticity. \textit{J. Econometrics} \textbf{31}(3), 307--327.
	
	\bibitem[\protect\citeauthoryear{Bougerol and Picard}{1992a}]{BP:1992a}
	Bougerol, P. and Picard, N. (1992a) Stationarity of GARCH processes and of some nonnegative time
	series. \textit{J. Econometrics} \textbf{52}(1-2), 115--127.
	
	\bibitem[\protect\citeauthoryear{Bougerol and Picard}{1992b}]{BP:1992b}
	Bougerol, P. and Picard, N. (1992b) Strict Stationarity of generalized autoregressive processes. \textit{Ann. Probability} \textbf{20}(4), 1714--1730.
	
%
%
	
	\bibitem[\protect\citeauthoryear{Chan and Ng}{2009}]{chng2009}
	Chan, N.H. and Ng, C.T. (2009). Statistical inference for non-stationary  GARCH$(p, q)$ models.
	\textit{Electron. J. Stat.} \textbf{3}, 956--992.
	
	
	\bibitem[\protect\citeauthoryear{Calzolari et al.}{2014}]{chp2014}
	Calzolari, G., Halbleib, R. and Parrini, A. (2014). Estimating GARCH-type models with symmetric stable innovations: Indirect inference versus maximum likelihood. \textit{Comput. Statist. and Data Analysis} \textbf{76}, 158--171.
	
	
	\bibitem[\protect\citeauthoryear{Chen and Wang}{2025}]{ChenWang2025}
	Chen, Y. and Wang, R., (2025). Infinite-mean models in risk management: Discussions and recent advances. \textit{Risk Sciences}, \textbf{1}(100003).
	
	\bibitem[\protect\citeauthoryear{Engle}{1982}]{Engle:1982}
	Engle, R.F. (1982). Autoregressive conditional heteroscedasticity with estimates of the variance of United Kingdom inflation. \textit{Econometrica} \textbf{50}(4), 987--1007.
	
	\bibitem[\protect\citeauthoryear{Fama}{1965}]{fama}
	Fama, E. (1965). The behavior of stock market prices.
	\textit{J. Bus.} \textbf{38}(1), 34--105.
	
	
	\bibitem[\protect\citeauthoryear{Fan and Yao}{2017}]{Fan:Yao}
	Fan, J. and Yao, Q. (2017). \textit{The Elements of Financial Econometrics.} Cambridge University
	Press, Cambridge.
	
	\bibitem[\protect\citeauthoryear{Francq and Zako\"{\i}an}{2004}]{FZ:2004}
	Francq, C. and Zako\"{\i}an, J.-M. (2004). Maximum likelihood estimation of pure GARCH and ARMA-GARCH processes. \textit{Bernoulli} \textbf{10}(4), 605--637.
	
	\bibitem[\protect\citeauthoryear{Francq and Zako\"{\i}an}{2012}]{FZ:2012}
	Francq, C. and Zako\"{\i}an, J.-M. (2012). Strict stationarity testing and estimation
	of explosive and stationary generalized autoregressive conditional heteroscedasticity models.
	\textit{Econometrica} \textbf{80}(2), 821--861.
	
	\bibitem[\protect\citeauthoryear{Francq and Zako\"{\i}an}{2013}]{FZ:2013}
	Francq, C. and Zako\"{\i}an, J.-M. (2013). Inference in nonstationary asymmetric GARCH models.
	{\it Ann. Stat.} {\bf 41}(4), 1970-1998.
	
	\bibitem[\protect\citeauthoryear{Francq and Zako\"{\i}an}{2019}]{FZ:2019}
	Francq, C. and Zako\"{\i}an, J.-M. (2019).
	\textit{GARCH Models: Structure, Statistical Inference and Financial Applications}, 2nd Edn. John Wiley.
	
	
	
	\bibitem[\protect\citeauthoryear{Hall and Yao}{2003}]{Hall:2003}
	Hall, P. and Yao, Q. (2003). Inference in ARCH and GARCH models with heavy-tailed errors. \textit{Econometrica} \textbf{71}(1), 285--317.
	
	\bibitem[\protect\citeauthoryear{Harvey}{2013}]{Harvey:2013}
	Harvey, A.C. (2013). \textit{Dynamic Models for Volatility and Heavy Tails:
		With Applications to Financial and Economic Time Series}. Cambridge University Press, Cambridge.
	
	
	
	
	\bibitem[\protect\citeauthoryear{Jensen and Rahbek}{2004a}]{jra}
	Jensen, S. T. and Rahbek, A. (2004a).  Asymptotic normality of the QMLE estimator of ARCH
	in the nonstationary case. \textit{Econometrica} \textbf{72}(2), 641--646.
	
	\bibitem[\protect\citeauthoryear{Jensen and Rahbek}{2004b}]{jrb}
	Jensen, S. T. and Rahbek, A. (2004b).  Asymptotic inference for nonstationary GARCH.
	\textit{Econometric Theory} \textbf{20}(6), 1203--1226.
	
	\bibitem[\protect\citeauthoryear{Khmaladze}{1981}]{Khmaladze}
	Khmaladze, E.V. (1981). Martingale approach in the theory of goodness-of-fit tests. \textit{Theory of Probability and its Applications} \textbf{26}(2), 240--257.
	
	
	
	
	
	\bibitem[\protect\citeauthoryear{Li, et.al.}{2023}]{Li:2023}
	Li, D., Tao, Y., Yang, Y. and Zhang, R. (2023). Maximum likelihood estimation for $\alpha$-stable double autoregressive models. \textit{J. Econometrics} \textbf{236}(1), 105471.
	
	\bibitem[\protect\citeauthoryear{Li and Zhu}{2020}]{Li:2020}
	Li, D. and Zhu, K. (2020). Inference for asymmetric exponentially weighted moving average models. \textit{Journal of Time Series Analysis}, \textbf{41}(1), 154--162.
	
	\bibitem[\protect\citeauthoryear{Liu and Brorsen}{1995}]{Liu:1995}
	Liu, S.M. and Brorsen, B.W. (1995). Maximum likelihood estimation of a GARCH-stable model. \textit{J. Applied Econometrics} \textbf{10}(3), 273--285.
	
	\bibitem[\protect\citeauthoryear{Mandelbrot}{1963}]{Mandel}
	Mandelbrot, B. (1963). The variation of certain speculative prices.
	\textit{J. Bus.} \textbf{36}(4), 394--419.
	
	\bibitem[\protect\citeauthoryear{Mandelbrot}{1997}]{Mandel1997}
	Mandelbrot, B. (1997). \textit{Fractals and Scaling in Finance: Discontinuity, Concentration, Risk}. Springer, New York.
	
	\bibitem[\protect\citeauthoryear{Matsui and Takemura}{2006}]{Matsui}
	Matsui, M. and Takemura, A. (2006). Some improvements in numerical evaluation of symmetric stable density and its derivatives. \textit{Comm. Statist. Theory Methods} \textbf{35}(1), 149--172.
	
	\bibitem[\protect\citeauthoryear{McCulloch}{1985}]{McCulloch}
	McCulloch, J. (1985). Interest-risk sensitive deposit insurance premia: Stable ACH estimates. \textit{Journal of Banking and Finance} \textbf{9}(1), 137--156.
	
	
	\bibitem[\protect\citeauthoryear{Mittnik et al.}{2002}]{Mittnik}
	Mittnik, S., Paolella, M. and Rachev, S. (2002). Stationarity of stable power-GARCH processes. \textit{J. Econometrics} \textbf{106}(1), 97--107.
	
	
	\bibitem[\protect\citeauthoryear{Nolan}{1997}]{Nolan1997}
	Nolan, J.P. (1997). Numerical calculation of stable densities and distribution functions.
	\textit{Comm. Statist. Stochastic Models.} \textbf{13}(4), 759--774.
	
	
	\bibitem[\protect\citeauthoryear{Nolan}{2020}]{Nolan}
	Nolan, J.P. (2020).
	\textit{Univariate Stable Distributions: Models for Heavy Tailed Data}.
	Springer, Cham.
	
	
	\bibitem[\protect\citeauthoryear{Panorska et al.}{1995}]{Panorska}
	Panorska, A., Mittnik, S. and Rachev, S. (1995). Stable GARCH models for financial time series. \textit{Applied Mathematics Letters} \textbf{8}(5), 33--37.
	
	%
%
%
	
	
	\bibitem[\protect\citeauthoryear{Samuelson}{1967}]{Samuelson}
	Samuelson, P.A. (1967). Efficient portfolio selection for Pareto-L\'{e}vy investments.
	\textit{The Journal of Financial and Quantitative Analysis} \textbf{2}(2), 107--122.
	
	
	
	
	\bibitem[\protect\citeauthoryear{Uchaikin and Zolotarev}{1999}]{UZ}
	Uchaikin, V.V. and Zolotarev, V.M. (1999).
	\textit{Chance and Stability: Stable Distributions and their Applications}.  VSP, Utrecht.
	
	\bibitem[\protect\citeauthoryear{Wang et al.}{2022}]{Wang2022}
	Wang, G., Zhu, K., Li, G. and Li, W.K. (2022). Hybrid quantile estimation for asymmetric power GARCH models. \textit{J. Econometrics}, \textbf{227}(1), 264--284.
	
	\bibitem[\protect\citeauthoryear{Zhang et al.}{2022}]{Zhang2022}
	Zhang, X., Zhang, R., Li, Y. and Ling, S. (2022). LADE-based inferences for autoregressive models with heavy-tailed G-GARCH(1, 1) noise. \textit{J. Econometrics}, \textbf{227}(1), 228--240.
	
	\bibitem[\protect\citeauthoryear{Zhu et al.}{2023}]{Zhu2023}
	Zhu, Q., Tan, S., Zheng, Y. and Li, G. (2023). Quantile autoregressive conditional heteroscedasticity. \textit{J. Royal Statistical Society Series B: Statistical Methodology}, \textbf{85}(4), 1099--1127.
	
	
\end{thebibliography}


\vskip .65cm
\noindent
Department of Statistics and Data Science,
Southern University of Science and Technology, Shenzhen 518055, China
\vskip 2pt
\noindent
E-mail: taoyx@sustech.edu.cn
\vskip 2pt

\noindent
College of Systems Engineering,
National University of Defense Technology, Changsha, China
\vskip 2pt
\noindent
E-mail: gonghuan@nudt.edu.cn
\vskip 2pt

\noindent
Department of Statistics and Data Science, 
Tsinghua University, Beijing 100084, China
\vskip 2pt
\noindent
E-mail: malidong@tsinghua.edu.cn

\end{document}